# Flexible Virtual Reality System for Neurorehabilitation and Quality of Life Improvement

**Iulia-Cristina Stanica** [1,2,*], **Florica Moldoveanu** [2], **Giovanni-Paul Portelli** [3], **Maria-Iuliana Dascalu** [1], **Alin Moldoveanu** [2] and **Mariana Georgiana Ristea** [4]

[1] Department of Engineering in Foreign Languages, University Politehnica of Bucharest, 060042 Bucharest, Romania; maria.dascalu@upb.ro

[2] Computer Science and Engineering Department, University Politehnica of Bucharest, 060042 Bucharest, Romania; florica.moldoveanu@cs.pub.ro (F.M.); alin.moldoveanu@cs.pub.ro (A.M.)

[3] Department 4, Carol Davila University of Medicine and Pharmacy, 050474 Bucharest, Romania; giovanni.portelli@drd.umfcd.ro

[4] Department of Dermatology, Elias University Emergency Hospital, 011461 Bucharest, Romania; mariana-georgiana.ristea@rez.umfcd.ro

* Correspondence: iulia.stanica@upb.ro; Tel.: +40-748795922

**Abstract:** As life expectancy is mostly increasing, the incidence of many neurological disorders is also constantly growing. For improving the physical functions affected by a neurological disorder, rehabilitation procedures are mandatory, and they must be performed regularly. Unfortunately, neurorehabilitation procedures have disadvantages in terms of costs, accessibility and a lack of therapists. This paper presents Immersive Neurorehabilitation Exercises Using Virtual Reality (INREX-VR), our innovative immersive neurorehabilitation system using virtual reality. The system is based on a thorough research methodology and is able to capture real-time user movements and evaluate joint mobility for both upper and lower limbs, record training sessions and save electromyography data. The use of the first-person perspective increases immersion, and the joint range of motion is calculated with the help of both the HTC Vive system and inverse kinematics principles applied on skeleton rigs. Tutorial exercises are demonstrated by a virtual therapist, as they were recorded with real-life physicians, and sessions can be monitored and configured through tele-medicine. Complex movements are practiced in gamified settings, encouraging self-improvement and competition. Finally, we proposed a training plan and preliminary tests which show promising results in terms of accuracy and user feedback. As future developments, we plan to improve the system's accuracy and investigate a wireless alternative based on neural networks.

## 1. Introduction

Neurological disorders affect more and more people nowadays, especially given the fact that life expectancy is currently increasing and we are facing an ageing population. According to the predictions of the United Nations, "by 2050, 1 in 6 people in the world will be over the age of 65, up from 1 in 11 in 2019" [1]. A highly comprehensive and detailed study spanned across three decades and including 195 countries shows the burden of neurological disorders [2]. In 2016, neurological disorders (including stroke, Parkinson's disease, central nervous system trauma and so on) represented the first cause of DALY (disability-adjusted life years) and the second cause of death (16.5% of global deaths) [2]. People suffering from neurological disorders have their quality of life (QoL) largely affected, in terms of physical activities, mental or social functioning (usually called generically as health-related quality of life (HRQL) [3]. Physical activities including walking, grabbing, clenching fist or other fine motor skills, day-to-day activities or deglutition are some of the functionalities commonly affected by neurological conditions that have serious consequences over the QoL of a person. In addition, cognitive, communication, or social functionalities are also affected. As neurological disorders represent the second cause of death at global level, innovative systems are needed for improving the QoL of the sufferers.

Specific physical exercises are beneficial for both rehabilitation procedures and for preventing the advancement of a disorder. As many neurological disorders are untreatable or have permanent effects, the rehabilitation should focus on treating the symptoms or reducing the disease's progression. Unfortunately, the number of therapists is limited, so that new alternatives are needed for globally improving the neurorehabilitation process [4]. The progress reached until now in the field of neurorehabilitation is not enough to reach the goals targeted by the United Nations in terms of sustainable development until 2030 [5]. More measures must be taken in order to help improve the quality of life of people affected by neurological diseases and decrease the effects or advancement of degenerative disorders. One of the ways of addressing the challenges faced by neurorehabilitation is by including consumable affordable devices based on emergent technologies (reducing the costs of each therapy session, increasing accessibility, reducing the number of therapists needed for face-to-face interventions by providing tele-rehabilitation capabilities). The evolution of virtual reality made it the main focus of numerous studies from the past decade which show that VR is a technology which facilitates the creation of realistic environments and training exercises suitable for a wide range of neurological disorders. Virtual reality can optimize individualized training, bring entertainment and motivation through gamified settings, replicate daily tasks and/or common environments (e.g., streets, shops etc.) and offer the perfect context for collecting the data needed for monitoring progress and evaluating the rehabilitation process [6].

We present Immersive Neurorehabilitation Exercises Using Virtual Reality (INREX-VR), a complex virtual reality (VR) system that facilitates the neurorehabilitation process, by allowing patients to self-train at home and have their progress monitored by specialists. The system targets neurological disorders such as stroke, Parkinson's disease and trauma of the central and peripheral nervous system, so that it includes both conditions with a very high prevalence (e.g., stroke) and those affecting people of all age ranges (e.g., Parkinson's affecting mostly elder people, the trauma of the nervous system affecting both young and elder) that can be rehabilitated using VR-based therapy. The flexibility of the proposed framework also makes it suitable for medical personnel training (including rehabilitation physicians, nurses, kinesiotherapists or medicine students). By using a variety of affordable consumer devices and sensors (HTC Vive with motion tracking, Myo gesture control armband including electromyography - EMG, and pulse measurement), a great number of exercises can be performed in order to help recover lost functionalities, delay the advancement of a degenerative disorders, or improve and maintain physical condition. Furthermore, the data provided by the sensors can be stored, shared and further analyzed by specialists, in order to assess the progress, the stage of the disorder and the training process. Although they are medically grounded, all proposed exercises are gamified, enabling a pleasant and engaging experience, maximizing patients' involvement, in order to ensure the constant, sustained training, for relevant results on people's QoL.

In the following chapters, we will present a study containing both an analysis of the current stage of neurological rehabilitation and our proposed approach. Section 2 presents the medical fundamentals of the neurological disorders chosen as a target for the system, and the existing solutions of classical therapy. Section 3 focuses on presenting a thorough research methodology and the state of the art of the modern, technology-based treatment methods. In Section 4, we describe the system, including functionalities, hardware and software components. Finally, Section 5 shows the preliminary results after testing the system with users, and we draw the conclusions and underline some possible future improvements in Section 6.

**2. Neurological Disorders and the Classical Treatment**

We will further present the neurological disorders targeted by our system in terms of causes, symptoms and their classical treatment.

Parkinson's disease (PD**)** is described as a progressive disorder which affects the nervous system, sometimes manifested through a slight tremble of one of the hands. The disorder's debut often passes unnoticed, as these mild symptoms specific to the disease's initial stage are usually ignored or associated to other causes (e.g., aging). Parkinson's disease has five main stages, where the patient's condition is degrading progressively: starting from mild symptoms which do not affect their quality of life (stage 1), reaching serious balance and movement issues (stage 3) and walking incapacity and need of permanent assistance (stage 5).

PD is an irreversible disorder of the nervous system, and since there is no existing treatment, prevention and diagnosis during early stages become extremely important.

Most of the symptoms of Parkinson's disease are related to movement. Tremors are present as precocious symptoms for 70% of PD patients; they affect firstly one finger, then they extend to the entire hand and finally to the lower limb. Tremor appears when the limb is in resting position, which in fact stands as one of the reasons why exercising is essential in the case of PD patients [7]. When the disorder is diagnosed by observing the upper limb tremor or the slowness of movements, the stage of the disease is rather advanced. The neurological manifestations of PD include: bradykinesia (slowness of movements), tremor of upper and lower limbs, high muscle tone, increased rigidity, with the back crooked similar to a question mark, freezing of gait (incapacity of moving even though the intention exists), lack of physiological co-ordination of arms when walking, micrographia (small and unreadable writing), hypomimia (facial expressions look like a mask), hypophonia (soft, quiet voice) [8].

Parkinson's disease has no cure, but specific treatments can slow it down or treat specific symptoms/improve lost functionalities. Some of the available treatments include medication, surgery [9], implanting pluripotent stem cells (a modern procedure based on personalized, patient-specific treatment, which uses fetal cell transplantation for completing the affected neurons) [10,11].

As movement is extremely affected in PD, physical activities become very important, especially during the first three stages of the disease. They have an important role and can have benefits over: walking, balance, fine motor skills, joint mobility, flexed posture, facial mimics [8]. Physical activity for treating Parkinson's disease must take into consideration the fact that the therapy must start with easy exercises, that do not exhaust the patient, and their intensity must increase gradually. Some recommendations include:

- Performing movements on the rhythm of the music, which can be both motivating and can facilitate the exercises;
- Applying guiding markers and limits (either visual, auditory or haptic)—this can be achieved with special glasses/headsets and sensors, which can help patients maintain the same pace and continue exercising;
- Staying hydrated (as neurons responsible for thirst are also affected);
- Perform physical exercises somewhere around noon (stiffness is increased in the morning, tiredness occurs in the evening).

During exercises, various health parameters must be monitored, including blood pressure, heart rate and oxygen saturation [12].

Stroke represents a neurological disorder which appears when the blood flow is severely diminished or even disrupted in a brain region. As a consequence, it produces the death of brain cells which are not irrigated and oxygenized. The time period of the neurons' destruction varies between a few minutes to several hours, producing instantly various neurological symptoms: paralysis, numbness or weakness of the face, upper or lower limbs (according to the body area controlled by the affected brain cells). The consequences of stroke can be severe, including disability or even death, therefore a quick medical intervention is essential for saving the patient's life and diminishing the disorder's effects and complications [13].

Stroke symptoms include: speech and understanding difficulties, muscle weakness or paralysis of a limb (upper or lower) or of the right/left side of the face, sudden headaches, dizziness, nausea, visual impairments, balance problems (falling, lack of coordination), unconsciousness, skin sensitivity of the affected limb/face side [13,14].

Stroke's general treatment requires the continuous monitoring of the neurological state, also including the measurements of important health parameters. Immediate interventions involve endovascular or surgical procedures. Under 1/3 of people affected by stroke manage to recover completely, and rigorous rehabilitation procedures are needed in order to maintain physical, intellectual and social functions in optimum parameters. Rehabilitation in the case of stroke is multidisciplinary, including specific medication, physiotherapy (intensive exercises involving the affected limb; if this is the case, functional electrical stimulation can improve efficiency), occupational therapy (for recovering work activities, daily activities, hobbies etc.), speech therapy, cognitive training for attention deficit.

For recovering the motor deficit, physical exercises can be performed either passively (with a therapist) or actively (alone), during the three broad stages of stroke rehabilitation [15,16]:

1. Flaccid stage—when the patient presents no muscle tone or voluntary movement, some of their reflexes disappear; correcting posture and applying therapeutic massage are used for maintaining joint and muscular integrity;
2. Spastic stage—after approximately 3 weeks since the occurrence of the stroke, muscle tone is increased, and voluntary movement control starts to appear; this stage includes kinetic rehabilitation programs, based on sensory-motor techniques, exercises for increasing muscular force and movement amplitude, as well as balance and posture; some of the used techniques are based on altering perception—mirror therapy (reflecting the movement of a non-affected limb over the affected one, creating the illusion of movement) [17], virtual reality therapy (virtual limbs created in the virtual environment, with real movements performed repeatedly, and usually augmented in the virtual world)—or on stimulation (muscular or cerebral stimulation).
3. Chronic/recovered stage—the motor deficit is established; therefore, it is very hard to recover at this stage; therapy is needed for health maintenance.

Our system focuses on helping people in the second rehabilitation stage but can be further used for health maintenance.

Other neurological disorders affecting the central and peripheral nervous system include spinal disc herniation, neuropathies, sciatica, carpal tunnel syndrome or trauma caused by work-related or car accidents. We will shortly describe two of the most common ones, disc herniation and neuropathies.

Disc herniation is a disorder which occurs when an intervertebral disc is deteriorated. The herniated disc can come in contact with or press on nerves or on the spinal cord, thus causing pain, numbness and weakness at the limb level [18].

In the case of disc herniation, there are several types of treatment available: conservative (avoiding body poses which are causing pain; exercising), medical treatment, kinesiotherapy, and surgery [19].

Neuropathies refer to a series of disorders which appear when nerves of the peripheral system (part of the nervous system, outside the brain and spinal cord) are being deteriorated. Different types of neuropathies include the affection of multiple individual nerves or the generalized affection of all peripheral nerves (e.g., neuropathy caused by diabetes). For 30% of cases, the cause of neuropathy is not known, 30% are caused by diabetes, while others include trauma, vitamin deficit, alcohol consumption, tumors, HIV and so on [20]. Present symptoms are motor or sensitive, including pain and limb numbness, and can lead to traumatic injuries, infections and metabolic disorders [9].

Neuropathies treatment can be done through medication, alternative treatment (chemical or electrical stimulation of the spinal cord), and physical exercises [9].

## 3. ICT Solutions for Neurological Disorders Treatment

To establish the current state of the research in this field, and for determining the functionalities, goals and technologies used in the INREX-VR system, we conducted a research methodology based on the systematic review of the available state of the art. We collected papers from four major databases: Elsevier (Scopus), IEEE Xplore, SpringerLink and PubMed, from January 1995 to June 2020. We used the Preferred Reporting Items for Systematic Reviews and Meta Analyses (PRISMA) [21] for studying relevant articles, for analyzing the feasibility of our system, establishing functionalities, targeted users and validating the technological alternatives (Figure 1).

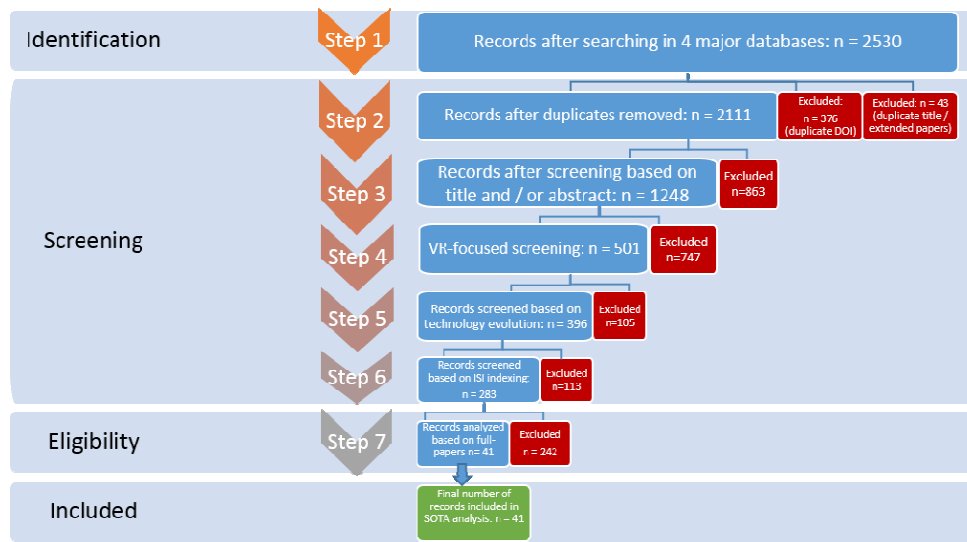

**Figure 1.** Preferred Reporting Items for Systematic Reviews and Meta Analyses (PRISMA) steps for the research methodology

The process of the PRISMA analysis that we applied includes four main stages (identification, screening, eligibility and inclusion), and seven detailed steps (as we divided the screening stage into four sub-stages).

In the first step, we managed to identify a total of **2530 papers from the four mentioned databases.** As a research tool, we used the following formula (1) for selecting the papers that contain these important topics in their title, abstract or keywords. The number of papers identified based on each database can be seen in Table 1:

$$\text{TITLE-ABS-KEY} = ((\text{"neurological disorder" OR "neurocognitive disorder" OR "neurorehabilitation" OR "neurological rehabilitation"}) \text{ AND "virtual reality"}) \quad (1)$$

**Table 1.** Number of papers identified in each scientific database.

| SpringerLink | PubMed | Elsevier (Scopus) | IEEE Xplore |
|---|---|---|---|
| 1129 | 750 | 564 | 87 |

In the second step, we started the screening process of the papers. We removed duplicates, by using two steps: firstly we removed the papers having the same digital object identifier—digital object identifier (DOI) (376 duplicates removed) and secondly, we compared the papers with no DOI based on a title/authors combination (thus eliminating duplicates or extended papers—43 additional duplicates removed). After this step, we obtained 2111 papers. At this moment, we could analyze the topicality of our research subject across the years (Figure 2). The increased interest can be easily observed in the last 20 years, coinciding with the technological advancement in the virtual reality field.

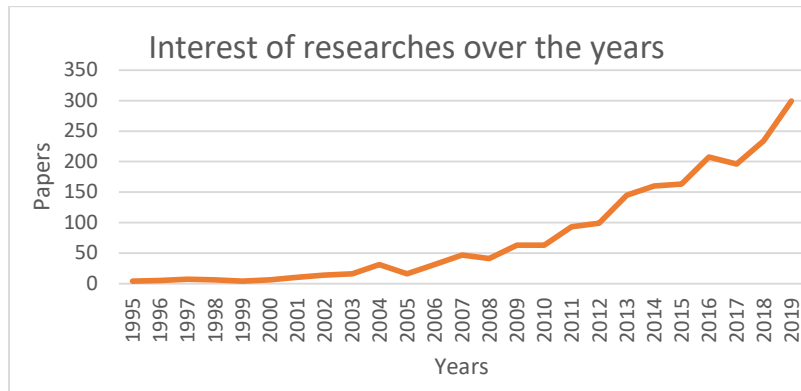

**Figure 2.** Evolution of the number of research papers related to virtual reality-based neurorehabilitation in the last 25 years.

In the third step, we continued the screening process by thoroughly analyzing the titles and abstracts of the selected studies; we thus eliminated 863 papers, including extended ones that were not focusing on our area of interest (e.g., books, chapters), papers that referred to other types of neurocognitive disorders that are not suitable for our research topic and rehabilitation system proposition (e.g., autism, dementia, anxiety disorders, phantom limb pain, cerebral palsy), papers focusing on economic aspects or legislation. We thus obtained 1248 articles.

After these screening steps, we analyzed statistics of the papers, to see what are the main technologies targeted. We identified papers focusing on robotics, virtual reality, augmented reality, mixed reality and reviews/surveys on one or more of these technologies used in rehabilitation scenarios, general platforms focusing on experiments and/or treatment (Table 2).

**Table 2.** Number of papers identified based on the main technology targeted or their type.

| Robotics | Virtual Reality | Augmented Reality | Mixed Reality | Reviews/Surveys | General |
|---|---|---|---|---|---|
| 296 | 593 | 17 | 8 | 166 (52/114) | 260 |

Since the work presented on this paper focuses on virtual reality, we eliminated the general papers and the ones focusing mostly on a different type of technology (robotics, augmented reality or mixed reality). Furthermore, we wanted to have a deep insight on the original projects using neurological rehabilitation and virtual reality, therefore, we eliminated also the reviews and the surveys. We thus performed the fourth step, the VR-focused screening and we got 501 papers (747 excluded).

For the fifth step, the screening was based on technological evolution**:** since technology evolves rapidly, including topics such as VR and sensors, we decided to keep only papers from the last 10 years (105 papers were removed) and we got 396 papers left from the period 2010−2020. For the sixth step**,** we decided to filter the papers based on their International Scientific Indexing (ISI)—therefore 113 papers were excluded and there were 283 left**.** In order to find the ISI papers, the advanced search was used for the entire batch of papers on Web of Science based on the papers' digital object identifier (DOI) or based on title and authors combination (if they had no DOI). After this last screening step, we analyzed the disorders targeted by the 283 papers remaining. In Table 3, we identified the number of papers concerning stroke, Parkinson's disease, brain trauma or spinal cord trauma. The other papers address either multiple disorders (e.g., central nervous system trauma in general) or focus on a certain functionality/therapy, regardless of the targeted disorder (e.g., upper limb therapy, gait or balance control). We can observe that a high number of papers are focusing on stroke (also probably related to the high number of people affected by this disorder and the prediction related to its future burden [22] and rise in younger people [23]). We must take this into consideration for designing our system, yet make sure it is flexible enough for a larger number of disorders.

Table 3. Number of papers identified based on targeted disorders.

| Stroke | Parkinson's Disease | Brain Trauma | Spinal Cord Trauma |
|---|---|---|---|
| 148 | 16 | 21 | 8 |

For the seventh and final step, we thoroughly analyzed the abstracts and the full papers (if the abstract was not detailed enough). We selected what we consider to be the most valuable for the final qualitative assessment and for establishing the characteristics of our system—therefore, scientific papers using sensors and trying to cover all three main disorders (PD, stroke, affections of central or peripheral nervous system) were preferred. In addition, we included all papers that presented home-based systems, since an important part of our proposed system has the focus on self-training. The final number of records that were included in our state of the art analysis was 41 papers. After the final step of the research, we concluded that all papers could be classified based on different criteria (Figure 3). We are going to focus for the rest of our state of the art analysis on the last classification type, based on technology.

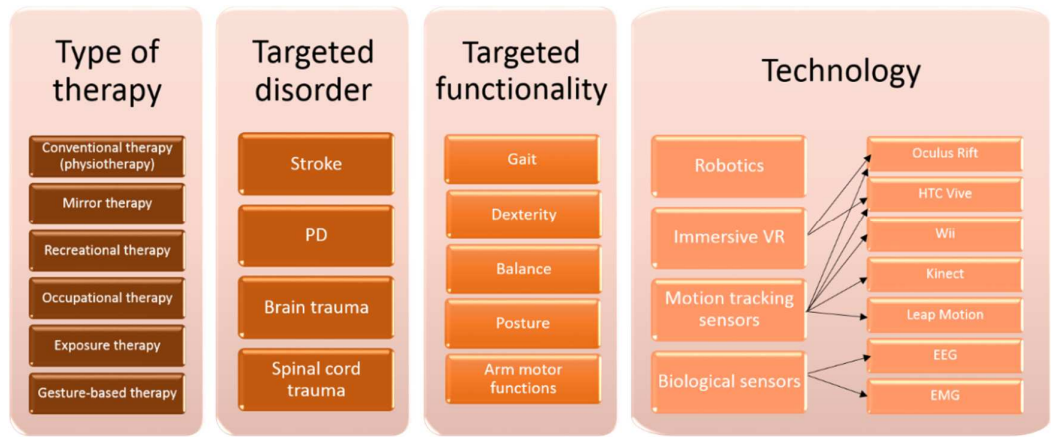

**Figure 3.** Classification of research papers of the state of the art analysis based on different criteria.

Among the first technologies used for neurorehabilitation purposes was robotics. From end-effector to exoskeletons, robotic devices have been used especially for upper and lower limb rehabilitation, with positive results [24,25]. Unfortunately, robot-based solutions are not suitable for home use, as they can be difficult to maintain, to store and can reach costs of several hundred thousands of dollars [26]. In clinical settings, combining robotics and VR can bring multimodal feedback, augment movement and speed up the recovery process [27]. Furthermore, studies show that combining these two technologies can bring great improvements for walking rehabilitation, as electroencephalogram (EEG) signals show that the brain activity is more intense when robot-assisted walking is accompanied by virtual environments (when compared with mirror therapy—seeing their own movements—or with unrelated visual stimuli) [28]. If robotics and VR-based neurorehabilitation also ensures the immersion of the person in the virtual environment [29] (e.g., by creating an accurate representation of the real limb in the virtual world and of the movements performed [30]), the user can become more motivated and encouraged to continue training. In addition, the evolution of sensors' reliability and accuracy of real-time adaptive learning algorithms can decrease the dependency of the patient on robotic or human assistance (from therapist or caregiver) [31].

Alternative technological solutions for neurorehabilitation include the use of virtual reality, motion and biological sensors, combined in different manners. Some of the earliest solutions from the last decade based on the extended definition of VR involved the use of a Nintendo Wii gaming console for upper-limb rehabilitation [32,33]. This gaming console is based on using various motion tracking sensors in a remote-control device for performing various movements (e.g., playing sports) with a virtual avatar that can be followed on a television screen, in a non-immersive virtual reality setting. Despite its lack of immersion, the Wii system has been proven to be more efficient for rehabilitation than other forms of treatment, such as recreational therapy [32], and as efficient as classical physical therapy [34]. Other versions of the device, such as Wii fit, include a standing board which can be used for balance training [35]. As many technical solutions focus only on upper body rehabilitation, Microsoft Kinect comes as an alternative of assuring full-body tracking [36]. Studies show that Kinect-based VR neurorehabilitation can provide low-cost solutions that can help brain reorganization which stands as the basis of motor recovery [37,38]. The principle of most Kinect applications is based on mapping human motions on poses of the virtual avatar's limbs, in a manner as accurate as possible [39], for instance when playing sports such as tennis or boxing [40].

Immersion plays a significant role in VR-based applications, as studies show that non-immersive systems can lead to lower performances when executing classical rehabilitation movements (e.g., flexion, extension, head rotation) [41]. Embodiment and sense of presence are proven to be improved for both healthy and neurologically affected individuals if the virtual training procedures are executed from a first person perspective as opposed to a third person one [42]. Furthermore, virtual mirror exercises (the user's movements are mapped on a virtual limb seen through an immersive VR headset) seem to generate a higher degree of corticospinal response when compared to real mirror therapy (seeing the reflection of an unaffected limb during training) [37]. Immersion can be ensured by the use of virtual reality headsets (e.g., Oculus Rift, HTC Vive, PlayStation VR). Contrary to some people's belief, the proper use of virtual reality headsets, with proper training and for reasonable time periods has no negative effects (headaches, falls or dizziness) even for persons affected by neurological disorders [43]. Furthermore, elder people with neurological disorders managed to use high-resolution head-mounted displays (HMD) without experiencing sickness or stress [44]. Therefore, Oculus Rift has been successfully used for head and neck rehabilitation [45] or for diminishing upper limb motor impairment [46]. Successfully combining motion detection systems (i.e., Kinect) with VR headsets (i.e., Oculus Rift) and using 3D game engines (i.e., Unity 3D) for creating gamified exercises can improve a person's mobility and quality of life [47]. In terms of high visual fidelity and the possibility of full-body tracking, HTC Vive systems come as a viable alternative, as studies show that their infrared-based motion tracking capability (for position and rotation) is extremely accurate (i.e., errors of "3.97 ± 3.37 mm" for human distance traveled, respectively, "0.46 ± 0.46 degrees" in angle rotation with a robotic arm) [48] when compared to high-cost motion capture systems [49]. As HTC Vive proves useful for rehabilitation, it has been used for lower-limb impairments in combination with motorized pedals [50].

Most of the papers selected for qualitative analysis show that biosensors (EEG, EMG, heart rate monitoring, etc.) are important in neurorehabilitation systems, for assessing different physiological parameters, such as brain or muscle activity during training. Electromyography (EMG) can be used for analyzing the degree of muscular activity when performing different VR exercises [51] or the use of myoelectric signals for controlling prosthetics [52]. VR rehabilitation game examples include the use of Microsoft Kinect and Myo gesture control armband (with EMG integrated) for treating people with neurological disorders [53]. Regarding EEG monitoring, this helps studying the effects of VR training over neural reorganization and can be used as a facilitator for a brain–computer interface (BCI) [54]. Studies show that BCI leads to better performances in neurorehabilitation systems if it is part of an immersive system [55].

Studies regarding the users' opinion related to VR neurorehabilitation systems underline that the key aspects for a successful system include "maintaining therapeutic principles", including "positive feedback" and "avoiding technology failure" [56]. Usual techniques developed for neurorehabilitation in VR systems include imitation, multiple repetition of the same exercises, eventually with different levels of intensity [57]. Augmentation is often used for enhancing the user's movements or reactions and impacting their motivation, either through physical means (e.g., pneumatic gloves [58], transcranial direct current stimulation [59] or functional electrical stimulation [60]) or through virtual ones (e.g., enhancing the person's movement in the virtual environment) [61]. In addition, systems tend to lean more and more towards home-use and tele-rehabilitation, as these can influence the person's motivation, well-being and desire to train, offering a more accessible solution regardless of physical impairments. Tele-rehabilitation systems allow therapists to carefully monitor the patient's progress and activity remotely, for both desktop [62] or mobile applications [63]. Tele-rehabilitation does not need to be seen as a replacement of rehabilitation centers, but rather as a tool which provides continuity once the patient is discharged [64].

To conclude, a virtual reality neurorehabilitation system should balance costs and efficiency, allow home-based training and be adaptable for a wide range of disabilities and situations. Furthermore, biosensor inclusion is mandatory, virtual environments and movements mapping should be immersive enough, while gamification principles must assure motivation and feedback.

## 4. Our System

After the thorough analysis of the existing state of the art regarding neurorehabilitation systems, we established the neurological disorders targeted by our own system proposal: stroke, Parkinson's disease and trauma of the central or peripheral nervous system. As for the main technology, we decided to use virtual reality. Rehabilitation treatment using virtual reality is based on many principles used in classical rehabilitation (e.g., performing exercises for joint mobility, executing daily tasks). The VR-based rehabilitation treatment is not considerably different in terms of rehabilitation procedures, and VR should be rather viewed as a facilitator for creating virtual environments, games, tasks, monitoring activity and collecting data, performing exercises at home, relieving the therapist's work by making them capable of supervising more patients at once through tele-rehabilitation.

Our neurorehabilitation system enables the creation of a wide range of customizable rehabilitation exercises which can be used for self-training using consumer-grade VR devices. Realism, embodiment, engagement, gamification and a sense of presence are essential for creating a motivating self-training environment. Gamification elements, such as score, badges, challenges and level of difficulties are present to raise the engagement of the user. Movement augmentation is obtained through amplifiers that help people apply less effort than normally required for performing a certain action (e.g., throwing a ball in a virtual scene). The goal is to help improve the quality of life of the people affected by the mentioned disorders by offering physical assistance and helping them improve lost or degenerating body functions and structures so that each person can become as independent as possible when performing daily activities. The real therapist is therefore represented in the virtual environment as a virtual avatar, while tasks and feedback are provided automatically by the system. The role of the real therapist is not neglected, as exercises and sensors data are centralized on a server from which they can be accessed and interpreted by specialists. The flexibility of the system also makes it suitable for medical personnel, such as medicine students, interns, nurses, and kinesiotherapists.

Furthermore, the software requirements were established based on the existing neurorehabilitation practices, procedures and protocols, also focusing on maintaining the suitability for elder people. Various virtual scenes and exercises are designed according to the model of the International Classification of Functioning, Disability and Health (ICF) [65], focusing on the interactions between different biopsychosocial factors [66] (p.10)—health condition, body functions and structure, activity, participation and contextual factors (environmental and personal).

Each of the designed exercises will be detailed according to the targeted biopsychosocial factors in the *Software components* sub-chapter.

In addition to patient suitability, ICF protocols are recommended for the training process of health professionals as well [67] (p. 46). They can be used for students' education and medical personnel training, providing a standardized framework based on a biopsychosocial approach. Therefore, we consider that a VR neurorehabilitation system based on ICF protocols can improve the education, collaboration and dedication of health professionals [68].

*4.1. System's Functionalities*

INREX-VR targets three main categories of actors: patients suffering from a neurological disorder, trainees (students, interns, residents, etc.) and therapists (responsible for supervising the rehabilitation of one or more patients). This system is suitable for people suffering from medium or low disabilities, not for those with advanced disabilities (e.g., bed bound). It can help sufferers with progressive neurological disorders so that they do not arrive in an advanced stage of gravity (e.g., suitable until stage 3 of Parkinson's disease) or those affected by neurological disorders so that they recover lost functions.

The main functionalities focus on patients' training (either through a "training tutorial" with classical rehabilitation movements, performed by a virtual trainer, or through rehabilitation games) and on medical personnel training (they can observe classical rehabilitation movements performed by the virtual avatar and practice them on their own). All users can register when they use the application for the first time and each patient will have their profile connected with their real-life therapist. Additional functionalities which target the therapist include the configuration of the training session of a patient (the therapist can view the profile of a patient and specify which of the system's exercises are suitable for the patient's condition and severity of disorder, as well as their levels of difficulty or level of augmentation for the movements) and evaluation of the session (the therapist can see the results of the exercises performed by the patient, in terms of accuracy and movements performed, as well as visualize and interpret the biosensors data). The emotional evaluation of the person during training includes heart-rate monitoring, level of stress through skin conductance or emotion detection through facial recognition, if the system is used without the virtual reality headset.

A simplified description of the system's main use cases is presented in Figure 4. Other external functionalities include the setup and calibration of the hardware devices and sensors, which can be done by the patients themselves in the case of at-home neurorehabilitation (if their health condition allows them and they have some technical knowledge), by a caregiver or a technical support specialist.

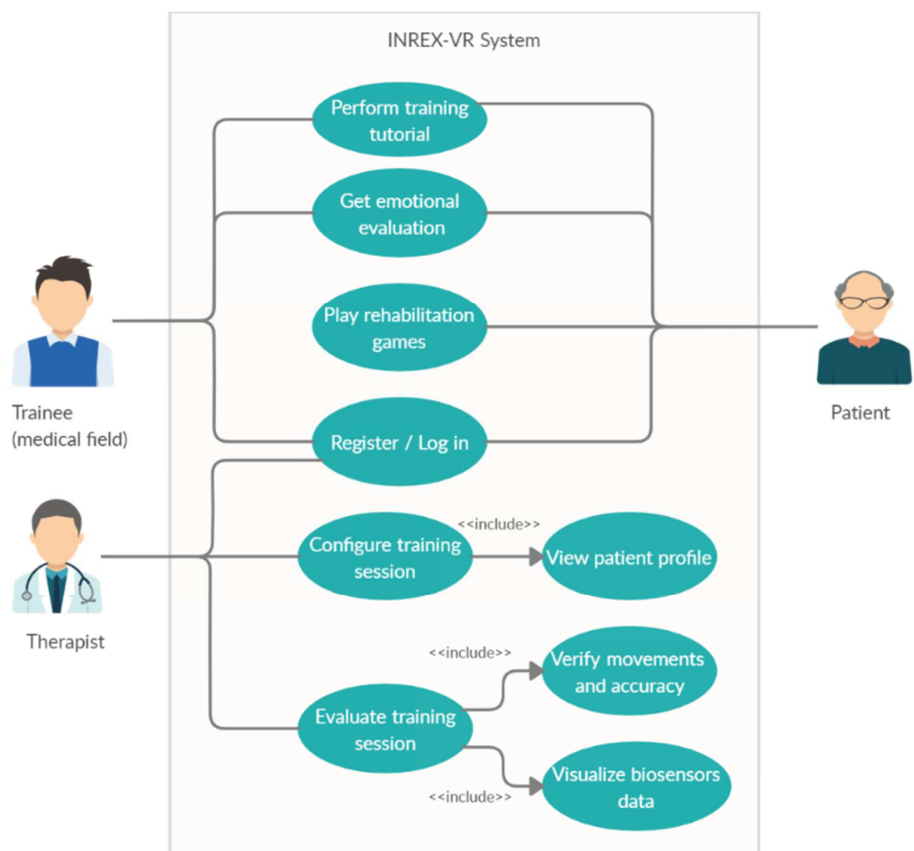

**Figure 4.** Use case diagram of the Immersive Neurorehabilitation Exercises Using Virtual Reality (INREX-VR) system.

*4.2. System's Architecture*

4.2.1. Hardware Components

The hardware components are chosen based on their technical performance, our previous experience with them and their suitability for our system's requirements. The complete physical architecture is displayed in Figure 5.

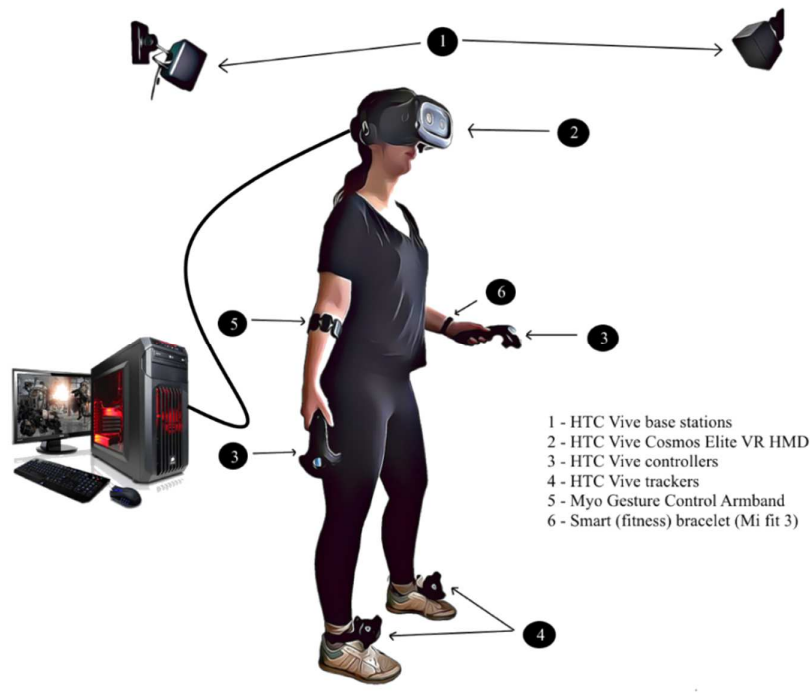

**Figure 5.** Hardware components of the INREX-VR system.

The central hardware component of the application is a complete HTC Vive Cosmos Elite VR system, including a headset, two HTC Vive base stations, two controllers and two Vive trackers. The HTC Vive Cosmos Elite headset offers a combined resolution of 2880 × 1770 pixels, representing an "88% resolution increase" when compared to the original Vive HMDs [69]. The HTC Vive system can be easily integrated in Unity3D (our main software development program) using the SteamVR software development kit. The VR system requires a first time configuration by placing the base stations in opposite corners and ensuring a minimum area of tracking of 2 m × 1.5 m [70]. Our standard setting includes the use of Vive controllers for arm movements and Vive trackers for leg movements in order to ensure the tracking of all limbs, but several alternative configurations have been tested: controllers for arm movements and one tracker in the torso region for full upper body tracking, trackers for arm movements instead of controllers for allowing the fingers' movement. Vive controllers ensure user input (through buttons, grip, trackpad), while both controllers and trackers can provide haptic output. The HTC Vive system must be connected to a high-performance PC. Our setup choice consisted of a high-end VR compatible PC (Intel Core™ i7-9700K 12 M Cache, up to 4.90 GHz processor, ASUS GeForce RTX 2070 ROG STRIX GAMING O8G 8 GB GDDR6 256-bit graphics card, 16 Gb DDR4 3200 MHz RAM memory and GIGABYTE Z390 GAMING X motherboard).

Biosensors analysis is essential in the field of neurorehabilitation, therefore we included another consumer device, the Myo gesture control armband. The device is placed on the user's forearm, and includes motion sensors for establishing orientation, rotation and acceleration, as well as eight EMG muscles sensors. Myo has several gestures configured by default (e.g., clenching fist, spreading fingers) and each patient can have their profile calibrated when using Myo for the first time, so that gesture recognition is mapped on their muscular profile. Myo must be connected via Bluetooth to the PC where the INREX-VR system is configured and can be integrated in Unity3D and Matlab with available software development kits (SDK). The armband can provide haptic feedback as well, with different levels of intensity.

As the patient's comfort and emotions must be monitored during the use of the INREX-VR system, we considered the integration of a Shimmer device for future tests, for monitoring the level of stress through skin conductance. For now, the heart rate is monitored using commercially available smart bracelets or smart watches. The Mi fit 3 fitness bracelet used in our system uses the photoplethysmography (PPG) technique to determine the heart rate and other health parameters (oxygen saturation, effort levels) with the help of proprietary algorithms. PPG is a popular technique among health monitoring devices, as it is inexpensive, non-invasive and often uses a green LED light source which emits light towards the person's skin. Green color is preferred as it reaches deeper through the tissue and a photodetector measures the amount of the reflected light to calculate the blood volume changes [71].

If the patient feels discomfort or is reluctant to use the VR headset, the INREX-VR offers also a lighter, limited version of use, as the gamified exercises targeting the upper limb are configured to work also just with Myo armband as a tracking device. In this case, emotion monitoring is ensured also by the emotion recognition software module which is described in Section 4.7. The simplified system for non-VR use is presented in Figure 6.

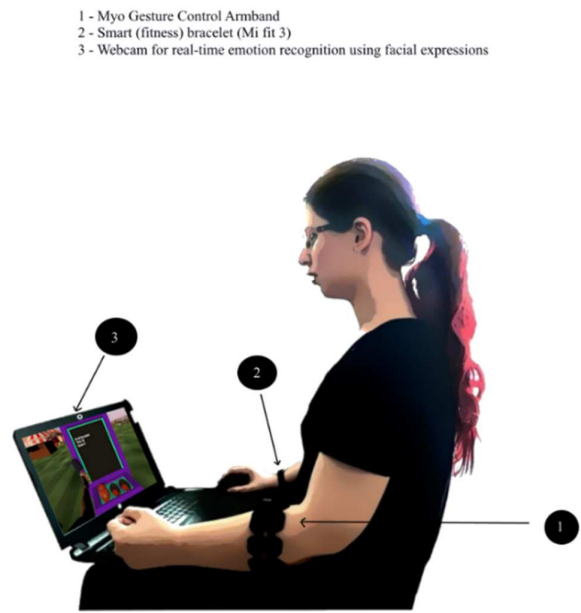

**Figure 6.** Simplified hardware setting for the non-VR use of the INREX-VR system.

4.2.2. Software Components

Virtual reality is be useful for the field of neurorehabilitation, as it can allow the creation of various rehabilitation exercises through 3D environments, ensure remote supervision from the therapist and encourage self-training, if it is sufficiently motivating. Motivation can be achieved through gamification, embodiment and usefulness of the designed scenes.

INREX-VR is a PC virtual reality application developed in C# using Unity3D game engine. The application is highly modularized, allowing its extension with new virtual scenes or neurorehabilitation exercises. The components of the application are displayed in Figure 7 and presented in more details in the following sections, including the external application programming interfaces (APIs) added to the system.

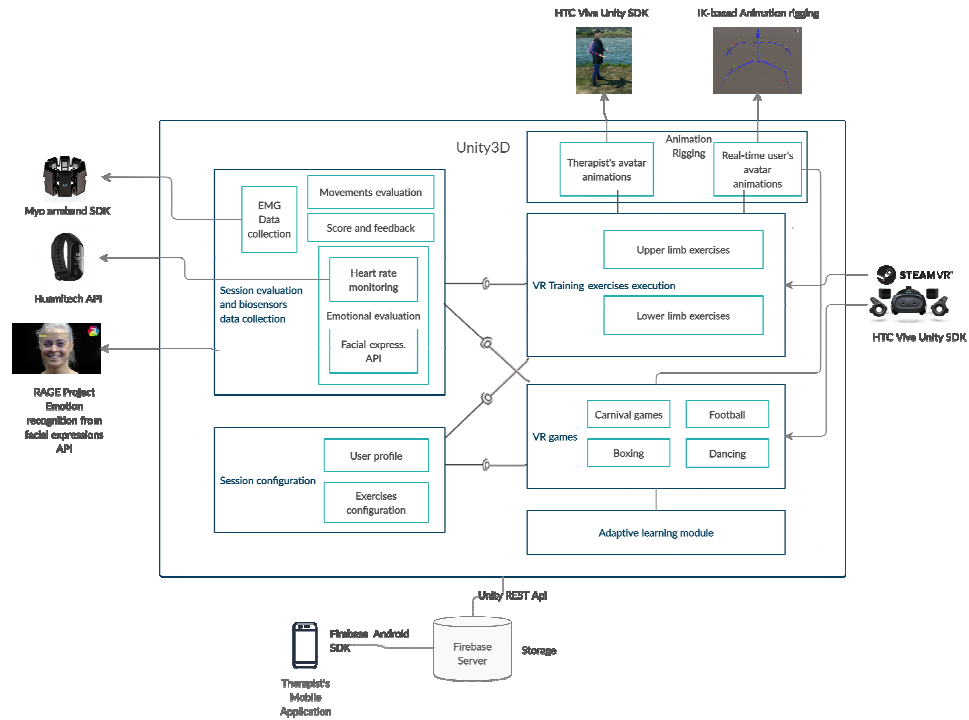

**Figure 7.** Software components of the INREX-VR system.

*4.3. Avatars Animations*

4.3.1. Real Time User's Avatar Animation

Studies analyzing the feeling of embodiment in VR environments date as far back as 1997 [72] and they underline the creation of a so-called "cyborg" connection—if the sense of presence is high enough in the virtual scene, the human mind can adapt more easily, creating a strong bond with the virtual environment. On the other hand, unnatural aspects in the virtual environment—unrealistic physics, unnatural body movement, inaccuracy between real gestures and those displayed in VR—can lead to detachment and lack of concentration. These aspects are essential for self-training VR systems, where people need to be engaged, stimulated and immersed in order to have a continuous, sustained training process.

The lack of consumer-grade full-body trackers has often led to compromises—VR games for instance usually show just the hand of the player rather than the whole upper limb in order to avoid any animation issue or decrease in embodiment. Many VR systems simulating body movements used expensive motion-capture sensors, but fortunately, the latest advancement in the field of consumer VR devices combined with inverse kinematics (IK) principles can lead to the successful creation of accessible and accurate simulations [73]. Inverse kinematics is highly used in robotics, from industrial robots to exoskeletons used in rehabilitation. This represents the process of calculating joint angles only from the coordinates of the end effector [74]. For instance, moving the hand will subsequently change the position and the rotation of the elbow and shoulder accordingly, in a natural manner. For achieving this with a VR avatar, the 3D model has to contain a skeleton which can be used for representing the kinematic chain from Figure 8.

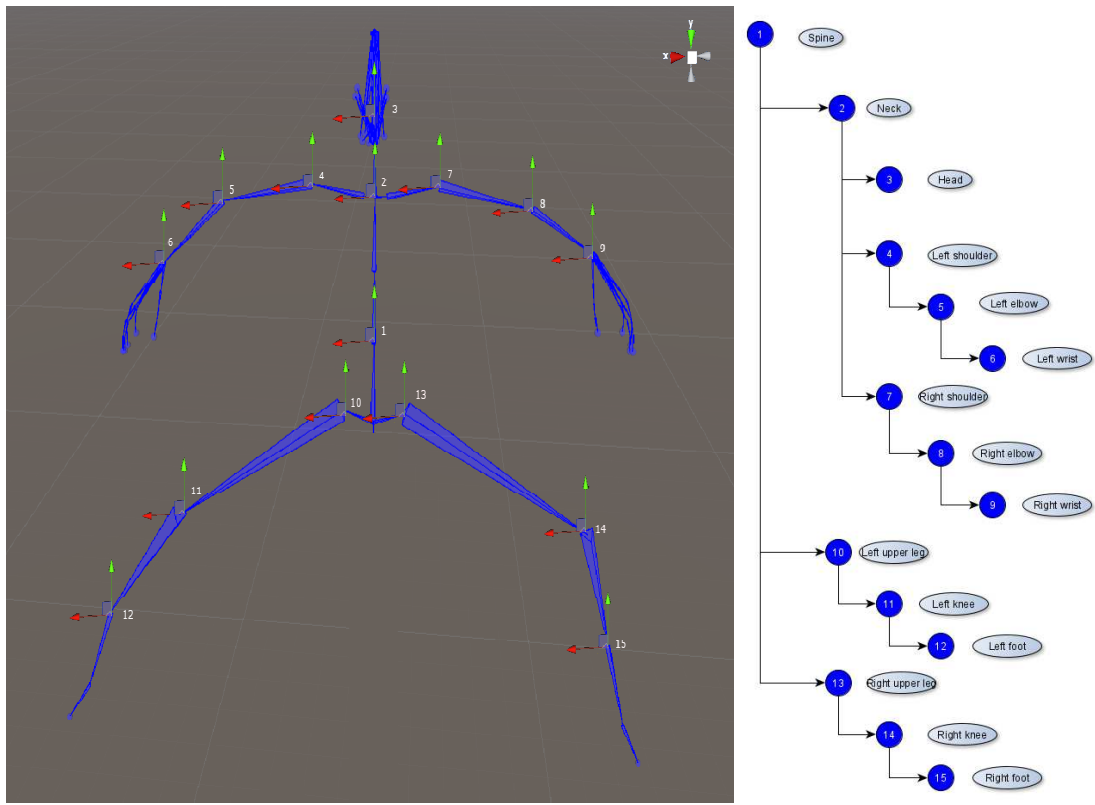

**Figure 8.** Avatar skeleton and relevant joints (**left**) and full-body kinematic chain (**right**).

In order to ensure the immersion and embodiment in the neurorehabilitation scenes, the user can visualize their virtual body from a first-person perspective. Using IK-based animations for the first-person avatar, the recovery person can see in their VR headset, in real time, the movements executed with their upper and lower limbs, or their head rotation. We achieve the inverse kinematics animations of the upper and lower limbs using Unity's prototype Animation Rigging tool [75]. We create the animations in a similar way for lower and upper limbs, based on the skeleton hierarchy displayed in Figure 8. Each animation of an individual limb is a rig based on a two-bone IK constraint—root bone (arm/upper leg), middle bone (forearm/lower leg)—controlled by the end-effectors (hand/foot). The appropriate transform pivots are placed in the two most important joints (e.g., wrist or ankle, where the sensors will be placed, and elbow or knee, the middle node used for computations). Basic IK principles are not sufficient, as the human body does not behave as robotic arms and additional constraints are needed (e.g., pronation and supination rotate both the wrist and forearm together, the wrist cannot be rotated individually with an angle higher than a threshold, etc.). Therefore, additional constraints were added for yaw, pitch and roll angles, based on testing basic gestures (circular feet motions, flexion/extension, deviations and so on), and we added twisting corrections for the elbow/hip joint to maintain natural movement. We also added animations to make sure that the body rotates and moves with the head, in order to ensure realism in every scenario (when our sensors are placed only on the head and limbs, and not on the torso). The final step was mapping the Vive sensors-based components to the rig effectors. This mapping includes the pairing of the camera-headset with the head pivot, controllers to the wrist pivots and trackers to the feet pivots. Offset values can be configured from the system's interface, for both position and rotation transforms, so that we can ensure the customization of the animations.

4.3.2. Therapist's Avatar Animations

When performing the tutorial rehabilitation exercises, the user must imitate the movements performed by the virtual therapist. We concluded that having an animated therapist avatar can increase motivation and entertainment for the user, therefore, we used the same hardware system for recording the exercises performed correctly by a real physician, which is part of the system's development team. With the help of the previously described IK system and the sensors attached on the real joints mapped on the therapist avatar's joints, we recorded all the transform changes (position and rotation) occurring at each frame in each one of the joints. All recorded frames are therefore combined in a single animation clip which can later be attached on any avatar with the same skeleton rig [76]. The correspondence between the real movements performed by the therapist and the resulting animation of the virtual avatar can be observed in Figure 9. The exercises are performed firstly with each limb at the time and at the end with both limbs together (in the case of the upper limb).

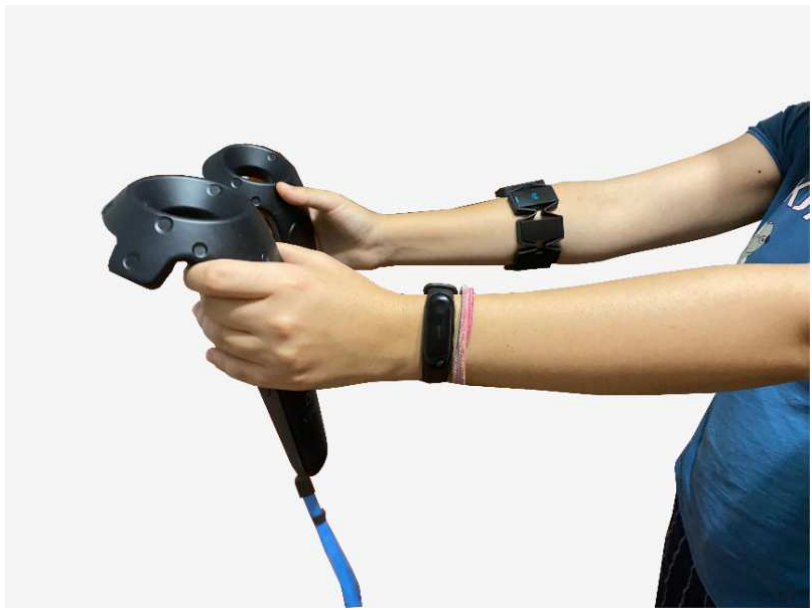

(**a**)

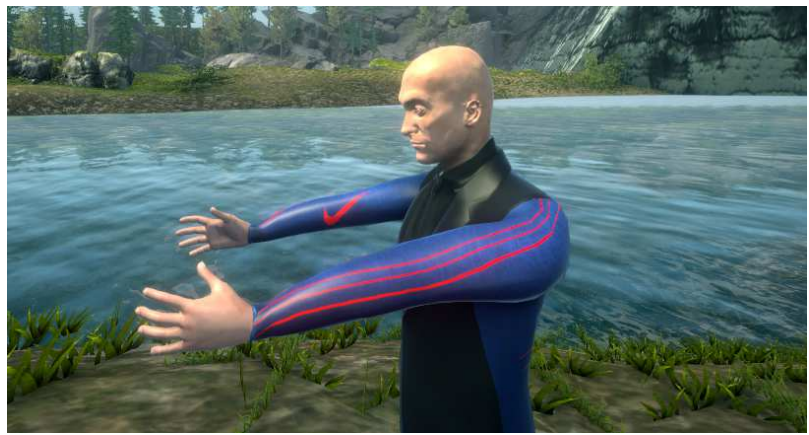

(**b**)

**Figure 9.** Rehabilitation exercise of elbow flexion (**a**) performed by therapist and (**b**) **the** corresponding animation of the virtual avatar**.**

All recorded tutorial exercises were chosen based on usual neurorehabilitation procedures applied for the targeted disorders [15] (pp. 29–54), in collaboration with physicians. We describe them more thoroughly in the following subchapter. Furthermore, the same recording module can be used by the patient during their own training sessions, if they want to see them later or send them to the therapist for visual reference.

*4.4. VR Training Exercises Execution*

As people are often new to VR as a technology, a tutorial is needed for helping them get used to the INREX-VR system. This is the first purpose of the VR training exercises module, as it contains basic exercises (for upper/lower limb), which are performed by the animated avatar and must be imitated by the user. The exercises were inspired by and are similar to classical exercises used by therapists in face-to-face neurorehabilitation processes, yet they are displayed in a virtual environment as executed by trainers and imitated by users through self-training. Each of the exercises is summarized in Table 4 according to their ICF biopsychosocial factors [65] and described in more details below. The environmental factors of all the exercises refer to the use of a technological product, for clinical or personal use.

**Table 4.** International Classification of Functioning, Disability and Health (ICF) biopsychosocial factors for each INREX-VR training tutorial exercise.

| Exercise | Health Condition(s) | Body Functions | Body Structures | Activities |
|---|---|---|---|---|
| Tutorial—upper limb | | | | |
| Shoulder flexion / extension 0°–90°, 90–180° Shoulder abduction / adduction | Stroke, Parkinson's disease, neuropathies | Mobility of joint, perception | Shoulder region, upper extremity (arm), brain, spinal cord and peripheral nerves | Learning and applying knowledge, communication, mobility |
| Arm pushing (downwards) Arm pushing (front) | Stroke, Parkinson's disease, neuropathies | Mobility of joint, muscle power, perception | Shoulder region, upper extremity (arm, hand), trunk, brain, spinal cord and peripheral nerves | Learning and applying knowledge, communication, mobility |
| Forearm extension / flexion Forearm pronation / supination Fist extension Fist adduction ("waving") | Stroke, Parkinson's disease, neuropathies | Mobility of joint, perception | Upper extremity (arm, hand), brain, spinal cord and peripheral nerves | Learning and applying knowledge, communication, mobility |
| Spinning wheel | Stroke, Parkinson's disease, neuropathies | Mobility of joint, orientation, perception, balance | Upper extremity (arm, hand), brain, spinal cord and peripheral nerves | Learning and applying knowledge, communication, mobility |

| Boxing training (jab punches) | Stroke, Parkinson's disease (early stages) | Mobility of joint, muscle power, muscle tone, perception, balance, energy and drive functions | Upper extremity (arm, hand), shoulder, brain, spinal cord and peripheral nerves | Learning and applying knowledge, communication, mobility |
|---|---|---|---|---|
| Tutorial—lower limb | | | | |
| Hip flexion Hip abduction Knee flexion (forward and backward) Ankle flexion/extension | Stroke, Parkinson's disease, disc herniation | Mobility of joint, perception, balance (if performed while standing) | Pelvis, lower extremity, brain, spinal cord and peripheral nerves | Learning and applying knowledge, communication, mobility |

The tutorial exercises can be performed in two distinct environments: a nature setting, where the avatar that exemplifies the exercises is a fitness coach, and a rehabilitation center, where the avatar is a physician. The implementation of the two environments has the purpose of verifying if a relaxed natural environment, with no obvious connection to a rehabilitation setting, can make the users more comfortable, motivated and dedicated during training [77].

All exercises must be performed as in a "mirror"—e.g., when the avatar executes the movement with their right arm, the user must execute it with their left arm. All exercises are shortly described based on the conventions used for the spatial position and the anatomic planes and sections [78], applied on our avatar's reference position (Figure 10). The values provided are corresponding to the highest degree of mobility of that particular joint (not affected) and will be used for the automatic evaluation of movements. It should be noted that the number of repetitions can be viewed more important than the precision, so people are encouraged to perform the exercises regardless of their mobility level. Furthermore, precision classes are important only for the testing procedure chapter, for data collection and research purposes only, if they are not displayed to the user during training. For the user, only information related to scores, time or challenges are important, as they can contribute to increasing their motivation and training will.

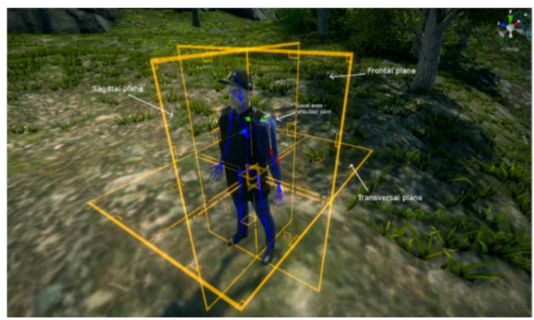

**Figure 10.** Anatomic planes and sections applied on the virtual avatar.

Shoulder joint exercises (full upper limb movements—Figure 11):

- Flexion 0°–90°, extension 90°–0° of the shoulder—starting position with the arms by the patient's side, palms facing downward (shoulder angle 0°); each arm should be lifted so that the shoulder joint is at 90° in the sagittal plane;
- Flexion 90°–180°/extension 180°–90°of the shoulder—second part of the previous exercise, on the segment 90°–180° in the sagittal plane (starting position with both arms in front, at 90°);
- Abduction 0°–90°/Adduction 90°–0° of the shoulder—starting position with arms hanging straight, in a natural way; each arm should be lifted so that the shoulder joint is at 90° in the frontal plane.

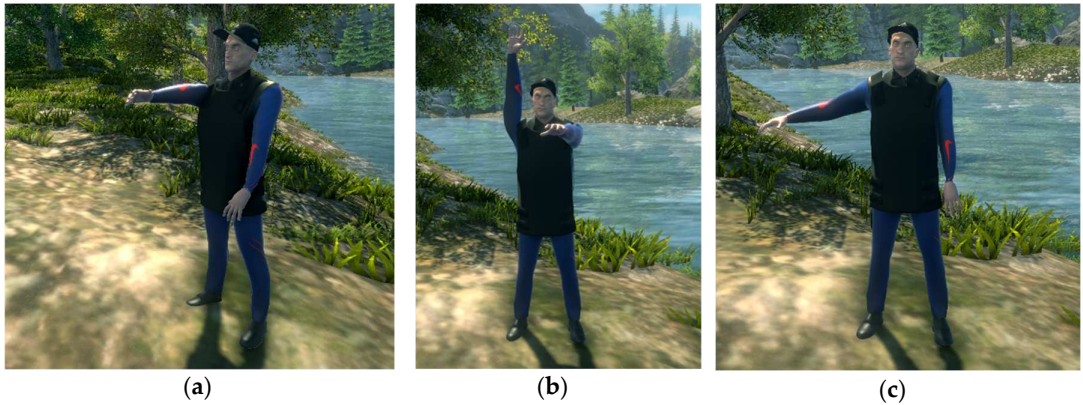

(**a**)　　　　　　　　　　(**b**)　　　　　　　　　　(**c**)

**Figure 11.** Shoulder joint exercises performed by the virtual avatar—left arm shows starting (resting) position and right arm shows the final position: (**a**) flexion 0°–90° of the shoulder; (**b**) flexion 90°–180° of the shoulder; and (**c**) shoulder abduction 0°–90°.

Elbow joint exercises—Figure 12:

- Extension/flexion of the forearm—starting position with the arms by the patient's side, palms facing upward (supination); the forearm should be flexed so that the elbow joint reaches at least 145° (up to 160°) (in the sagittal plane). The extension is represented by bringing the forearm to the initial position (elbow angle 0°);
- Pronation/supination of the forearm—supination and pronation are rotational movements in the transversal plane (along the longitudinal axis of the limb), supination representing the upwards movement of the palm, and pronation the downward movement [15] (p. 41). The movement is performed from 0° to 90° (supination), respectively, from 0° to −90° (pronation), the neutral position being the one with the arms oriented and fully stretched forward, with the thumb oriented upwards;
- Arm pushing (downwards)—training both elbow and fist joints; from a relaxed position, with the hands close to the chest and the forearm oriented horizontally so that the elbow joint forms an angle of approximately 90° in flexion, a pushing motion is made towards the ground until the forearm is fully extended (elbow joint angle becomes 0°);
- Arm pushing (front)—training both elbow and fist joints; from a relaxed position, with hands raised at the shoulder level and the forearm oriented vertically (hands close to face), a forward pushing movement is performed until the forearm is perfectly extended (elbow joint angle becomes 0°), and the fist joint is at least 60° in extension in the sagittal plane.

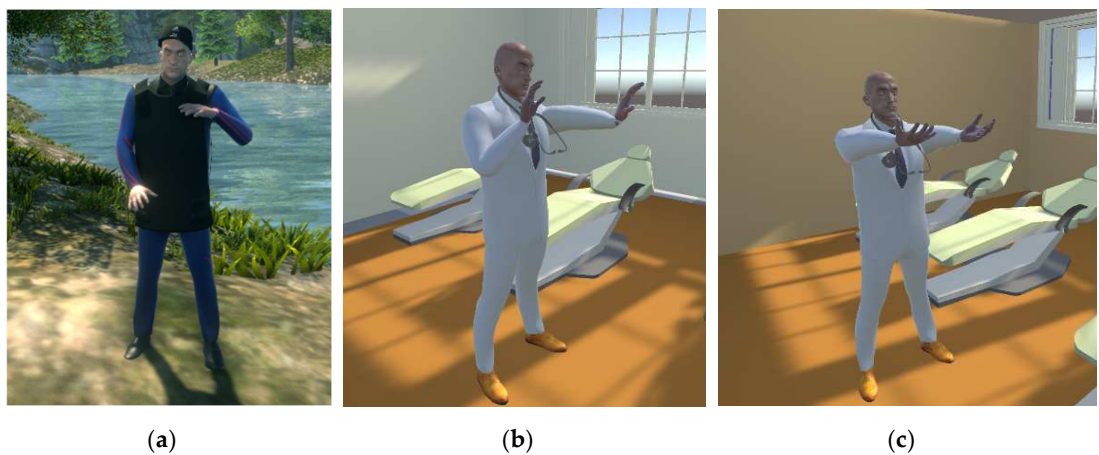

(**a**)　　　　　　　　　　(**b**)　　　　　　　　　　(**c**)

**Figure 12.** Elbow and fist joints exercises performed by the virtual avatar: (**a**) arm pushing (downwards); (**b**) arm pushing (front); (**c**) pronation/supination of the forearm.

Fist joint exercises:

- Fist extension (Figure 13a)—with the arms stretched forward, from a pronation position, the fist joint must be rotated from 0° to 70° in the sagittal plane;
- Fist adduction (Figure 13b)—similar with the waving gesture; from the extension position presented in the previous exercise, the joint must be adducted with a maximum of 50°–55° in the frontal plane (exterior rotation).

General upper limb exercises (force, coordination):

- Spinning wheel—classical exercise especially for Parkinson's disease, where the movement coordination is tested. The initial position is the same from the "arm pushing (downwards)" presented previously. The symmetry of the execution of the movements is evaluated (the successive rotation transforms of the arm joints must have comparable values), so that the joint of each elbow forms an angle of at least 45° in the sagittal plane throughout the movement; the physical resistance is measured (for how long can the user execute the exercise without losing focus or coordination);
- Boxing training (jab punches) (Figure 13c)—this exercise aims to train the user for the boxing game (Section 4.5), with the most basic movement of this sport—the jab punch. The starting position is similar to that of the exercise entitled "arm pushing (front)", but the hand is relaxed or with a clenched fist, not in extension. The user must make a pushing movement towards an imaginary adversary, at chest or face level, with each arm at the time, until the forearm is perfectly extended; when one arm is pushed forward, the other returns to the neutral position ("guard"); the physical resistance (the time of repetition of the hits without interruption) and the speed of the hits are evaluated.

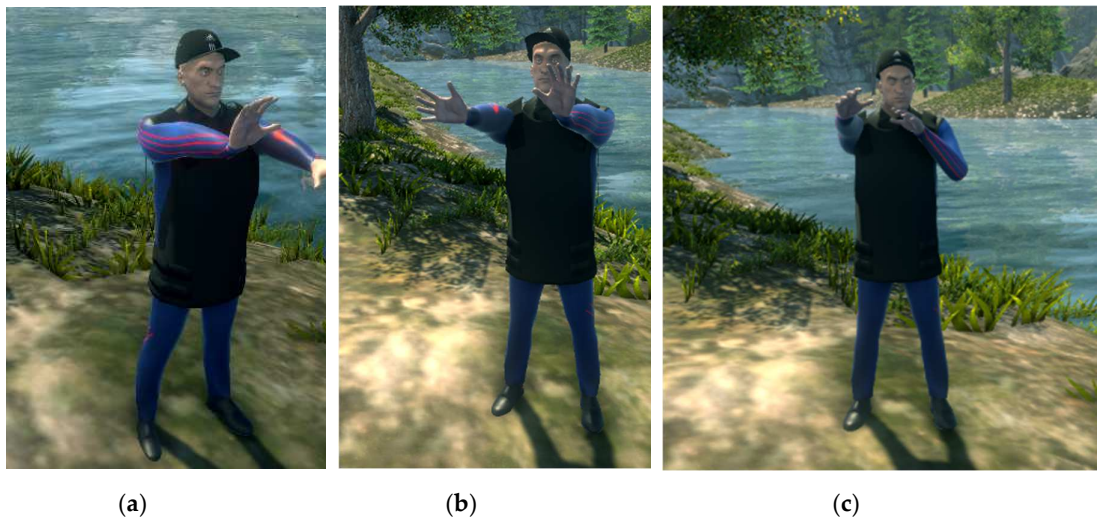

(**a**)　　　　　　　　　(**b**)　　　　　　　　　(**c**)

**Figure 13.** Fist joint and general upper limb exercises performed by the virtual avatar: (**a**) fist extension; (**b**) fist adduction; and (**c**) boxing training (jab punches).

Exercises for the lower limb (Figure 14) are exemplified by the trainer standing up, but most of them can be performed by the patient when lying down (on the back or on the side), if their physical condition does not allow them to stand up.

Hip joint exercises (whole lower limb):

- Hip flexion (0–90°)—the lower limb should be raised forward, keeping it perfectly stretched so that the hip joint approaches 90° in the sagittal plane;
- Hip abduction—the lower limb must be raised to the side, keeping it perfectly stretched so that the hip joint approaches 45° in the frontal plane; the patient's pelvis must remain still and not tilt to the opposite side.

Knee joint exercises:

- Knee flexion (forward)—the lower limb should be raised forward, with the knee flexed in the sagittal plane;
- Knee flexion (backward)—the previous exercise can be executed also with the lower limb oriented backwards; both movements should consist of at least 75°–80° knee rotation in the sagittal plane.

Ankle joint exercises:

- Ankle flexion/extension—similar to a "toe stand" followed by "heel stand"; standing on toes is equivalent to an extension of about 45° in the sagittal plane; the flexion is represented by returning to the initial position, with a slight bend on the heels (maximum 20°).

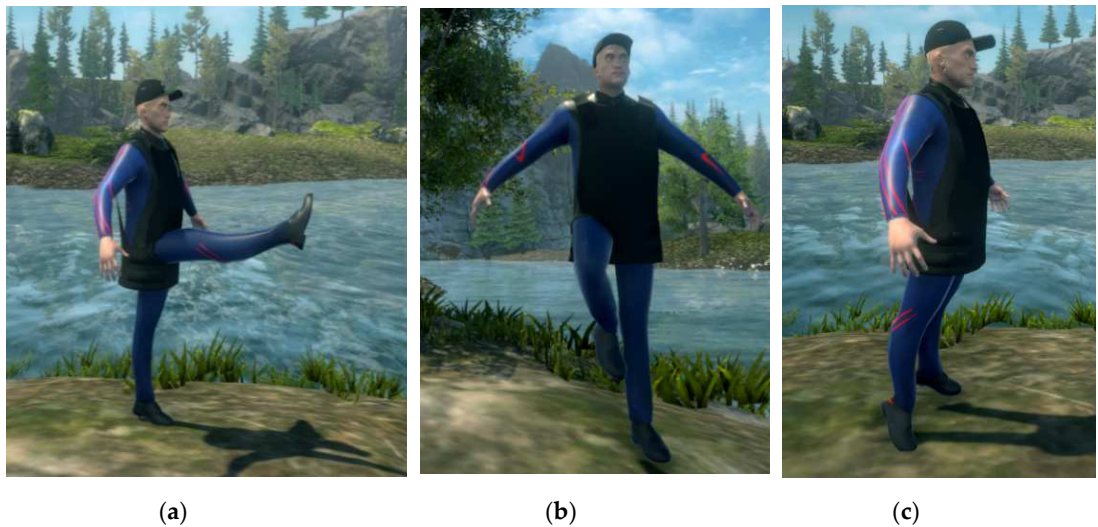

(**a**)　　　　　　　　　　　　(**b**)　　　　　　　　　　　　(**c**)

**Figure 14.** Lower limb exercises performed by the virtual avatar: (**a**) hip flexion (0°–90°); (**b**) knee flexion (forward); (**c**) ankle extension.

*4.5. Training in VR Games*

The second part of the neurorehabilitation process of a patient involves training in different VR games. The users must follow the verbal or written indications and apply the moves learnt in the tutorial for accomplishing gamified tasks. These activities focus on adding fun, gamified elements (score, levels of difficulty, challenges and badges), for motivating the patient to continue the rehabilitation process. All VR games are assessed based on their ICF characteristics, as presented in Table 5. Furthermore, the gamified exercises help the user train complex skills needed for daily activities and require the same movements: e.g., picking up a ball and raising it to launch it requires the same action as picking up a glass to drink.

The gamified rehabilitation exercises are providing multimodal feedback: visual (virtual reality environments where real-world movements are mapped accordingly), haptic (impulses provided by the hardware devices for simulating physical contact in the games, applied on the targeted limb) and auditory (music, verbal tasks and hints, jingles).

Table 5. ICF biopsychosocial factors for each INREX-VR game.

| Exercise | Health Condition(s) | Body Functions | Body Structures | Activities |
|---|---|---|---|---|
| Carnival games | | | | |
| Hit targets Ball directing | Stroke, Parkinson's disease, neuropathies | Muscle power, orientation, attention, perception | Shoulder region, upper extremity (arm), brain, spinal cord and peripheral nerves | Applying knowledge, undertaking simple and multiple tasks, mobility, communication |
| Whack-a-mole | Stroke, Parkinson's disease, neuropathies | Mobility of joint, movements, orientation, attention, perception | Shoulder region, upper extremity (arm), brain, spinal cord and peripheral nerves | Applying knowledge, undertaking simple and multiple tasks, mobility, communication |
| Boxing | | | | |
| Guard, multiple, complex hits | Stroke, Parkinson's disease (early stages), neuropathies | Mobility of joint, muscle power, muscle tone, perception, balance, energy and drive functions | Upper extremity (arm, hand), shoulder, brain, spinal cord and peripheral nerves | Applying knowledge, undertaking simple and multiple tasks, mobility, communication |
| Lower body games | | | | |
| Football Dancing | Stroke, Parkinson's disease, disc herniation | Mobility of joint, muscle power, muscle tone, attention, perception, energy and drive, balance, orientation (if performed while standing) | Pelvis, lower extremity, brain, spinal cord and peripheral nerves | Applying knowledge, undertaking simple and multiple tasks, mobility, communication |

Upper limb exercises

The upper limb exercises include carnival games (where therapeutic movements are performed in gamified settings) and boxing, great for both rehabilitation and maintaining good physical shape.

- Hit targets—picking up a ball and throwing it to hit a tower of cans. The exercises of shoulder flexion–extension (0°–180°) from the tutorial are practiced now, as well as actions of grabbing/releasing of objects. The actions are performed in a very natural manner. Clenching the fist will activate the controller's grab buttons, respectively, Myo's specific gesture, for picking up a ball; the user must lift their arm and launch the ball at a certain speed; by releasing the grab buttons/unclenching the fist, the ball will be launched. Various levels of difficulty are obtained through different a number of hit targets or different distances between the player and the targets. Picking up the ball is simulated through vibrations (from the controller or the armband).
- Ball directing—a ball must be directed on a table which has a ramp at its end and land in holes with different scores. This game practices movements of the elbow from the tutorial, and a new ball will be generated in the person's hand, which is the grip buttons of the controller being pressed. Similar launching actions are performed when the grip buttons are released. Difficulty is varied through the maximum number of available balls or through the distance between the player and the table's end. Picking up the ball is simulated through vibrations (from the controller or the armband).

- Whack-a-mole—the user must hit as many moles as they can using a hammer, in 60 s. Both hand and elbow tutorial movements are being trained. The HTC Vive scenario includes the use of a 3 × 3 matrix of moles. Various difficulty levels are obtained by changing the moles' spawning frequency and the duration until they are "hiding" back in their holes. Hitting a mole is simulated through vibrations (from the controller or the armband).
- Boxing—the user must perform different boxing techniques (guard, jab, cross punch, hook, uppercut), as performed by a virtual trainer. There are two scenarios—in a boxing ring with a mannequin and in front of a punching bag (heavier than the mannequin). The score is calculated based on the number of punches thrown in one minute. The contact with the target is simulated through vibrations (from the controller or the armband).

The upper limb exercises are also implemented to be used in a simplified manner without the HTC Vive system (in special situations when the user feels uncomfortable when using the headset or if the person only has access to limited hardware devices from the system). In this situation, the setting is adapted to be used with the Myo armband and actions are triggered based on the gestures performed with it (clenching fist for picking up objects, spreading fingers for releasing them, left only right movement of arm for hitting left–right targets and so on). The difference in perspective for the non-VR and VR scenarios, respectively, can be seen in Figure 15. For the whack-a-mole exercise, there are only three moles in the non-VR scenario, as the armband cannot track depth movements of the arm, so the moles are displayed on one row and must be hit by performing left–right translations.

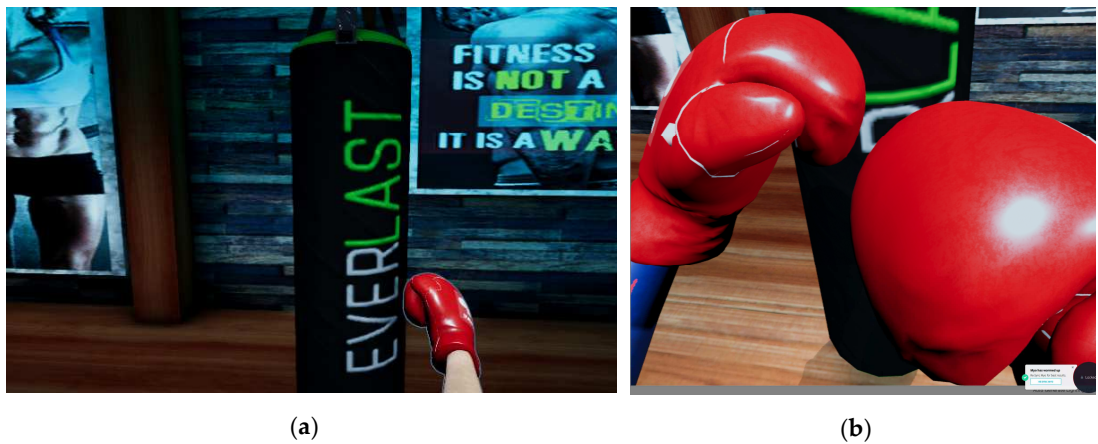

(**a**)        (**b**)

**Figure 15.** Carnival games: (**a**) perspective on screen in non-VR scenario; and (**b**) perspective through the VR headset.

Lower limb exercises

Initially, the gamified rehabilitation exercises developed in our system with the animated avatar integrated included basic physiotherapy movements (for upper/lower limb) and games such as picking up a ball and throwing it to hit targets, ball directing games, coordination games (whack-a-mole), boxing, or slowly walking in virtual environments. However, walking in VR can cause motion sickness without proper training or for people who are new to this technology, and preliminary tests with people not familiar to using VR showed that they feel discomfort and/or dizziness when the walking action from the real world is mapped similarly into the virtual one. This is why we changed our lower limb exercises and games, so that they focus more on the flexion/extension of the limb (e.g., football, dancing game—both done with the person standing in one fixed spot) rather than walking for long distances.

- Football—ball shooting to hit the goal from a fixed position. All lower limb joints are being trained. Various degrees of difficulty include varying the distance to the goal (e.g., penalty, free kick), with or without a goalkeeper or a wall and hitting from a central or lateral position. Hitting the ball is simulated through vibrations of the tracker.
- Dancing—similar to the "arcade dancing games". All lower limb joints are being trained. The user must touch colorful squares on the floor that are being lit in the rhythm of the music, as shown on an arcade screen. Different songs need different speeds of performing the steps and vary the difficulty.

Because realism and attention to detail can have a major impact on user motivation, efficiency and dedication, the scenes were designed with great attention to the graphics, avoiding elements that can diminish the user's immersion. For example, when using the Myo armband without the HTC Vive system (and as a result seeing only the avatar's arms, without the rest of the body), a separate camera is used for permanently rendering the arms, as they should not pass through walls or other non-relevant objects that the user should not collide with (e.g., tables, arcade machines). The lighting of the static objects is realized in Unity's "baked" mode, to be calculated before runtime and not dynamically in each frame (thus improving performance, which is essential for a VR application). Improved graphic effects are obtained by post-processing (Figure 16)—effects such as vignette, fog, ambient occlusion, depth of field, and color changes on RGB channels, with some of them deactivated in the case of VR (e.g., the vignette is visually pleasant on computer screens but can be tiring for the eyes on a headset).

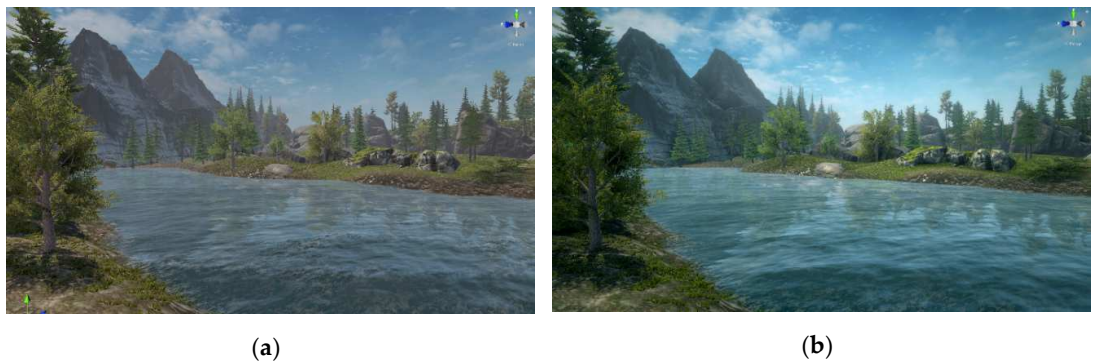

(**a**)　　　　　　　　　　　　　　　　　　　(**b**)

**Figure 16.** Nature scene graphics: (**a**) before post-processing; (**b**) after post-processing.

*4.6. Adaptive Learning Algorithm*

Adaptive learning is essential in a self-training application or distance learning system [79]. For keeping people motivated and encouraging them to self-improve, we decided that the games should include an adaptive learning algorithm which includes all gamified exercises. Each game starts with a neutral task (medium level of difficulty) and based on the user's performance, the adaptive algorithm will determine the next level which should be performed by the user. Adaptive elements include varying the number of targets or the distance between the user and the targets (hitting cans, directing balls), spawning frequency and hitting time window (whack-a-mole), ball position—centered or on a side—and distance to the goal (football), force required (all exercises) or demos shown by a trainer (boxing).

The adaptive learning concept is summarized in Figure 17, including the three main levels of difficulty and variations which can appear at each individual level, with examples taken from the games. The current version of the adaptive learning algorithm takes only one parameter in consideration (e.g., distance) and will be further developed to include all aspects mentioned in this subchapter.

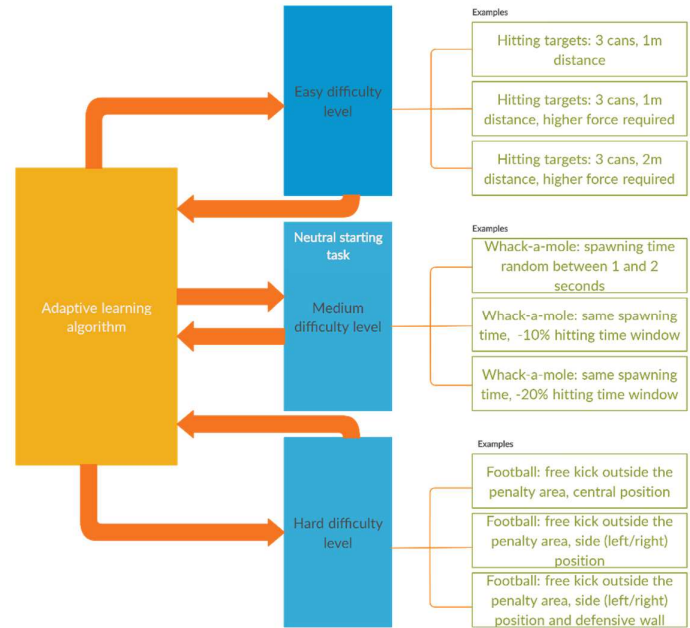

**Figure 17.** Adaptive learning algorithm concept schema.

*4.7. Session Configuration, Evaluation and Data Collection*

For starting a training session without a configuration from the therapist (i.e., if the system is used for the training of the medical personnel), tutorial exercises or games can be selected from the menu where they are grouped based on their targeted limb and/or joint. Session configuration can be personalized for the user's profile, by selecting the most suitable exercises for their disorder or degree of impairment either from the Unity application (useful for clinical use or medical training) or from a mobile application. Different parameters can be customized, such as the number of repetitions to be performed, the training duration, as well as the force that the user must apply for succeeding in a game (this parameter is controlled through a force amplifier which must have greater values for more serious impairments, thus assuring movement augmentation). If the therapist recommends performing only one specific level of difficulty, they can manually disable the adaptive learning algorithm and set their desired level.

In addition to getting used to the system, each exercise in the tutorial has the role of automatically evaluating the degree of mobility at the level of each joint. Thus, in each exercise, the maximum angle reached after performing the movement is analyzed, respectively the position of the relevant limb subdivision, based on mobility coefficients. The movements recorded by the therapist based on specialized rehabilitation treaties [15] are considered to have the ideal values accepted and the system calculates the deviation of the movements performed by the user from these standard values. We therefore obtain a score of the exercise and can approximate the functional deficit of one or more joints. Each position/rotation evaluation is done according to the local coordinate system of each joint, respecting the previously mentioned conventions related to the spatial position and the orientation planes of the human body (e.g., for the left shoulder joint as shown in Figure 18, the axes are represented as follows: Ox—red, Oy—green and Oz—blue; an abduction/adduction movement of the shoulder, i.e., lateral lifting arms, will require a rotation around the Oy axis). Other data are also automatically computed by the system, such as the number of hits/score, the time of execution, and the speed of movement/force.

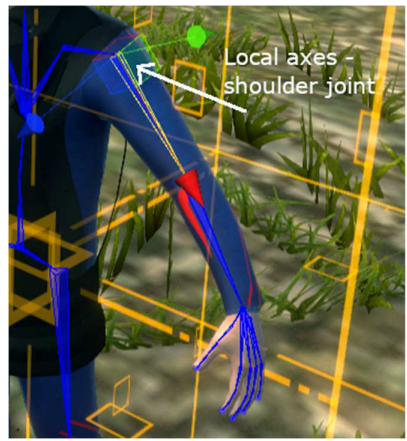

**Figure 18.** Axes in the shoulder joint (local coordinate system).

Relevant data, including game parameters and performance, movements and biosensors data are being saved locally and sent to a Firebase Realtime Database server through a Unity REST API [80]. They can be accessed from the same server by the therapist through their mobile application, in order to see the patient's results and progress and change their exercises' configuration accordingly. Emotion-related data are also saved in the same database (heart-rate values or emotions, in the case of non-VR use of the system). For detecting emotions through the facial expressions captured by the webcam, we used the RAGE (Realising an Applied Gaming "Emotion Detection Asset" created by the Open University of the Netherlands [81] which is capable of identifying the six basic emotions: happiness, sadness, surprise, fear, disgust, and anger [82] or if the user is having a neutral expression.

## 5. Preliminary Results and Discussion

### 5.1. Testing Procedure

A preliminary laboratory testing study was conducted as a validation procedure, including six healthy subjects (Users 1–6). All healthy subjects from the first stage of testing were involved in the creation of the project as part of the development team or as physicians consulted for designing the exercises and establishing the functionalities of the system. According to clinical trials procedures, it was mentioned by the World Health Organization and the American National Institute of Health that testing a new drug, vaccine or medical device needs an "initial safety evaluation in healthy adults" [83]. These healthy volunteers helped collect the data needed in studies for comparison, also facilitating the development of new knowledge or the improvement of system functionalities [84]. Besides the definition of "normal" parameters (e.g., scores in rehabilitation games), tests with healthy subjects help assess the possible risks caused by virtual reality so that we can establish risk management strategies before testing the system with patients (e.g., side effects caused by virtual reality, a lack of comfort of devices, a lack of relevance or appropriateness of exercises) [85]. The second step of the testing studies included two elder patients suffering from neurological disorders, who tested a simplified version of the INREX-VR system (without the HTC Vive), testing all games for the upper limb and all training exercises for both upper and lower limb. These two users are further referred as User 7 and User 8. Our future step of testing, after performing all improvements underlined from these two preliminary tests, includes the use of the complete VR system with a higher number of patients. Unfortunately, the advancement of the testing phase is impossible at the moment, as we were not able to have access to patients in hospitals or rehabilitation centers in the past 6 months because of the COVID-19 pandemic restrictions in our country.

All test subjects provided their informed consent (Supplementary Materials S1_Patient-consent-form). before starting the study and all protocols from the Declaration of Helsinki were respected. The overview of the test subjects can be observed in Table 6, including their self-expressed levels of technological and VR experience (skills) before testing the INREX-VR system, based on a five-point

Likert scale (i.e., excellent, good, fair, low, very low). The level of physical activity is appreciated based on the "Saltin–Grimby Physical Activity Level Scale" (inactive, light physical activity, regular physical activity, hard physical activity, such as for competitive sports) [86].

**Table 6.** Overview of the test subjects.

| User ID | Age | Activity Level | Technological Skills | VR Skills |
| --- | --- | --- | --- | --- |
| User 1 | 25 | Light activity | Good | Very low |
| User 2 | 29 | Regular activity | Good | Fair |
| User 3 | 27 | Hard activity | Excellent | Excellent |
| User 4 | 55 | Regular activity | Low | Low |
| User 5 | 59 | Inactive | Very low | Very low |
| User 6 | 55 | Light activity | Fair | Low |
| User 7 | 84 | Inactive | Very low | Very low |
| User 8 | 87 | Inactive | Very low | Very low |

The VR system setup presented in Section 4.2.1 was used for all healthy participants (Users 1–6). As VR components, User 4 used only the HTC Vive headset and upper limb controllers, without leg trackers. The HTC Vive headset was connected with the PC via cable, but the person assisting the testing procedures made sure it did not have any negative impact on the user's experience (as the cable and testing zones were spacious enough). A high-end configuration allowed us to present the system to the user with the highest graphics quality settings of the 3D scenes without any latency issues, allowing us the recording process of some training procedures at the same time. The tests were conducted under the same conditions, with only one exception: when the health condition did not allow them to perform the exercises while standing up, the users were allowed to perform all exercises sitting down. Otherwise, the conditions are the same for all VR test subjects: same laboratory setting (same room, surface, light conditions, temperature), same software exercises mentioned in the procedure, same VR setup (headset, controllers for hand and trackers for feet). For the two users suffering from neurological disorders, the testing took place at the users' personal residence, with the non-VR version of the system (laptop and Myo gesture control armband, plus emotional evaluation from the facial expressions using the webcam).

The testing procedure consisted of four main phases: phase 0 (accommodation—filling consent form, measuring health parameters, presenting, configuring and getting used to the system), phase 1 (training exercises—one or two representative exercises for each one of the joints, performed one at a time, saving the maximum joint rotation values in text files and recording the training sequences as an animations clip so that they can be re-watched and analyzed in detail); phase 2 (games—four upper limb games and one lower limb performed for 3 min, either adaptively in one long trial or with predefined levels of difficulties, in three distinct trials; scores and system mentions saved in text files); phase 3 (final feedback—answering a questionnaire with different aspects related to the use of the system). During the training phases of the procedure (phases 1 and 2 below), parameters such as the heart rate and EMG were continuously monitored. EMG datasets from the patients are grouped based on the exercises performed and are available as Supplementary Materials (S2_CompleteUserTestingData). Table 7 describes in detail each one of the phases in terms of purpose, activities and performance measurements.

Table 7. Detailed testing plan.

| Phase ID | Purpose | Activities | Performance Measurements | Duration |
|---|---|---|---|---|
| Phase 0 | System presentation and accommodation | - Informed consent<br>- Initial health parameters measured (heart rate, blood pressure)<br>- System presentation (hardware and software)<br>- VR accommodation (if applicable)<br>- User muscular profile calibrated on the Myo gesture control armband<br>- VR system configuration according to the user's data (if applicable) | - User feedback | One hour |
| Phase 1 | Classical training exercises for both upper and lower limb<br><br>Goniometer and system's recorded values comparison | - Fitness bracelet exercising mode started<br>- Upper limb exercises, 5 repetitions/trials with each limb at a time and 5 repetitions with both at the same time:<br>• Shoulder: flexion 0°–90°, abduction 0°–90°;<br>• Elbow: forearm flexion, supination;<br>• Fist: arm pushing, extension.<br>- Lower limb exercises, 5 repetitions/trials with each limb at a time:<br>• Hip: flexion, abduction;<br>• Knee: flexion;<br>• Ankle: flexion. | - Joint angles (system) for each individual trial<br>- Goniometer angle for first trial<br>- System accuracy—per exercise, per patient, overall<br>- Average joint angle for right/left limb across all 5 trials<br>- Mobility degree (according to mobility classes established in Annex 1)<br>- Average execution times for each exercise | One hour |

| Phase 2 | Gamified training | - Presentation of input for performing in-game actions and accommodation time (a few minutes) for each game<br>- Hit targets: 3 min adaptively, from easy and gradually increasing difficulty levels (medium, difficult)<br>- Ball directing: 3 trials of one minute each, the user must beat their previous record<br>- Whack-a-mole: 3 trials of one minute each with different levels of difficulty (easy, medium, difficult)<br>- Boxing—3 trials of one minute each in different settings (2 in the ring—easy, 1 with the punching bag—medium)<br>- Football—3 trials with 12 shots each (2 from penalty distance, 1 from free kick distance) | - Score according to each game's logic<br>- Performance classes of each game (according to the classes established in Annex 2)<br>- Hit targets: maximum difficult reached, number of cans hit in each hit, number of tries to complete a level<br>- Ball directing game: number of holes hit<br>- Whack-a-mole: number of moles hit, maximum number of moles that could have been hit, accuracy<br>- Boxing: number of hits with right and left fist; applied force<br>- Football: number of goals, accuracy | 30–45 min |
| Phase 3 | Final feedback | - Feedback collected related to topics such as:<br>• Opinion on exercises, scenes, graphics<br>• VR perception (negative effects, discomfort, embodiment)<br>• Suggestions for improvement | - User feedback | 15–20 min |

*5.2. Synthetic Results*

For each one of the training exercises from phase 1, we calculated the average angle across all trials for each user's right/left limb, respectively, and we evaluated the range of motion degree (ROM) of each limb, as well as the execution time of each exercise (for both right and left limb together), for the upper body (Table 8) and lower body (Table 9), respectively. The range of motion degree evaluation was established with the help of physicians, based on mobility coefficients used in rehabilitation treaties [15] (pp. 33–34). We thus established a five-point Likert scale which is further described in Appendix A, Table A1 and includes the following marks: excellent (5), good (4), fair (3), poor (2), very poor (1). For personal health reasons, User 4 only executed the upper limb exercises which were performed while sitting down. Individual data collected at each trial for each user can be found in the Supplementary Materials (S2_CompleteUserTestingData, Tables *Table_User1* – *Table_User8*).

According to the statistics, users performed best in the shoulder category, obtaining very good and good ROM results. For the fist exercises, the results were the lowest, partially because holding the Vive controllers limits the hand's range of motion to a certain extent and prevents finger extension. Further tests are required using also trackers for hands, similar to the ones used for feet. The use of trackers can also solve the elbow supination abnormality—normally, the movement is restricted to 90° by musculoskeletal reasons, and values that are higher (e.g., User 2 and User 5 in Table 8) are obtained by involuntarily rotating the controller independently from the fist joint (i.e., rotating the device but keeping the hand still). From the average values table, as well as the detailed trial recordings (Supplementary Materials S3_RecordingAnimations), we can observe that some exercises were performed at a higher angle than required (e.g., User 4 has an average angle of 109.4° for the right arm, and 108.6° for the left arm in the shoulder flexion exercise, even though they were asked to perform a 90° movement). In this particular case, a higher angle was translated as a very good level of mobility, but the lack of proprioception can be caused by the fact of not being accustomed with the first person perspective from a virtual world. The user must be therefore helped and messages should be displayed on screen when a movement was correctly performed.

**Table 8.** Upper limb mobility statistics for each healthy user and average results across all healthy users. All corresponding scores for the ROM are mentioned in Table A1.

| User ID | Shoulder Flexion | | | Shoulder Abduction | | | Elbow Flexion | | | Elbow Supination | | | Arm Pushing | | | Fist Extension | | |
|---|---|---|---|---|---|---|---|---|---|---|---|---|---|---|---|---|---|---|
| | Avg. Angle Right/Left | ROM Right/Left | t(s) Right + Left | Avg. Angle Right/Left | ROM Right/Left | t(s) Right + Left | Avg. Angle Right/Left | ROM Right/Left | t(s) Right + Left | Avg. Angle Right/Left | ROM Right/Left | t(s) Right + Left | Avg. Angle Right/Left | ROM Right/Left | t(s) Right + Left | Avg. Angle Right/Left | ROM Right/Left | t(s) Right + Left |
| User 1 | 96.4 / 89.5 | 5 / 5 | 60.5 | 82.4 / 83.6 | 5 / 5 | 50 | 123.2 / 119.6 | 5 / 5 | 52.5 | 45.4 / 57.8 | 3 / 4 | 81 | 60.4 / 59.4 | 5 / 4 | 90 | 46.2 / 35.6 | 3 / 3 | 52 |
| User 2 | 88.8 / 85 | 5 / 5 | 79.5 | 105.2 / 103.2 | 5 / 5 | 61 | 151.8 / 136.4 | 5 / 5 | 70.5 | 105.8 / 102.6 | 5 / 5 | 138.5 | 59.4 / 49.4 | 4 / 4 | 81 | 63.6 / 62.8 | 4 / 4 | 74 |
| User 3 | 86.6 / 92.8 | 5 / 5 | 80 | 90.6 / 88.8 | 5 / 5 | 89 | 115.2 / 116.4 | 5 / 5 | 69 | 70.8 / 71.2 | 4 / 4 | 98 | 34 / 54.2 | 3 / 4 | 97 | 35.8 / 36.2 | 3 / 3 | 52 |
| User 4 | 109.4 / 108.6 | 5 / 5 | 117 | 90.2 / 87.4 | 5 / 5 | 97 | 99.2 / 94.2 | 4 / 4 | 136 | 80.2 / 91.4 | 5 / 5 | 132 | 66.4 / 62.6 | 5 / 5 | 75 | 67.4 / 67.8 | 4 / 4 | 50 |
| User 5 | 94.8 / 98.4 | 5 / 5 | 118 | 95.2 / 93.4 | 5 / 5 | 59 | 125.4 / 114 | 5 / 5 | 76 | 99.6 / 102.4 | 5 / 5 | 128 | 53.4 / 48.2 | 4 / 4 | 107 | 71.2 / 52.2 | 5 / 4 | 44 |
| User 6 | 87.4 / 88.6 | 5 / 5 | 98 | 80.8 / 78 | 5 / 4 | 87 | 120.4 / 107 | 5 / 5 | 82 | 86.6 / 82.2 | 5 / 5 | 105 | 39.6 / 45 | 3 / 4 | 99 | 57 / 48.4 | 4 / 3 | 62 |
| Avg. (Right/Left) | 93.9 / 93.81 | 5 / 5 | 92.2 | 90.73 / 89.07 | 5 / 4.83 | 73.8 | 122.53 / 114.6 | 4.83 / 4.83 | 81 | 81.4 / 84.6 | 3.67 / 4.67 | 113.7 | 52.2 / 53.13 | 4 / 4.17 | 91.5 | 56.87 / 50.5 | 3.83 / 3.5 | 55.7 |

Table 9. Lower limb mobility statistics for each healthy user and average results across all healthy users. All corresponding scores for the ROM are mentioned in **Table A1.**

| User ID * | Hip Flexion | | | Hip Abduction | | | Knee Flexion | | | Ankle Flexion | | |
|---|---|---|---|---|---|---|---|---|---|---|---|---|
| | Avg. Angle Right/left | ROM Right/Left | t(s) Right + Left | Avg. Angle Right/Left | ROM Right/Left | t(s) Right + Left | Avg. Angle Right/Left | ROM Right/Left | t(s) Right + Left | Avg. Angle Right/Left | ROM Right/Left | t(s) Right + Left |
| User 1 | 88.4 / 85.6 | 5 / 5 | 32.5 | 65.4 / 57.2 | 5 / 5 | 42.5 | 109 / 130.2 | 5 / 5 | 37.5 | 21.4 / 26.4 | 4 / 4 | 69.5 |
| User 2 | 74 / 54.6 | 4 / 4 | 70 | 53.4 / 40.6 | 5 / 4 | 109.5 | 70.8 / 55 | 4 / 4 | 66.5 | 34.6 / 36.6 | 4 / 5 | 94.5 |
| User 3 | 98.8 / 99.4 | 5 / 5 | 92 | 60.2 / 72.8 | 5 / 5 | 60 | 122.6 / 130.6 | 5 / 5 | 125 | 25.6 / 25.6 | 4 / 4 | 32 |
| User 5 | 67.4 / 68.4 | 4 / 4 | 52 | 46.8 / 50.8 | 5 / 5 | 45 | 98.8 / 107.4 | 5 / 5 | 43 | 49 / 31.4 | 5 / 4 | 58 |
| User 6 | 86.2 / 89.4 | 5 / 5 | 84 | 67.2 / 64.2 | 5 / 5 | 59 | 83.4 / 80.4 | 4 / 4 | 106 | 36.2 / 40.4 | 5 / 5 | 62 |
| **avg. (right/left)** | 82.96 / 79.48 | 4.6 / 4.6 | 66.1 | 58.6 / 57.12 | 5 / 4.8 | 63.2 | 96.92 / 100.72 | 4.6 / 4.6 | 75.6 | 33.36 / 32.08 | 4.4 / 4.4 | 63.2 |

\* User 4 performed only upper limb exercises for personal health reasons.

Table 9 shows that all users obtained very good and good results across all lower limb exercises. The performance is overall better than the one obtained for upper limb exercises, showing that Vive trackers are more suitable as they are kept in a fixed position on the user's foot (shoe) and do not restrict the movements in any way. The ankle flexion exercise obtained the lowest performance scores, partially because users were asked to have the trackers fixed on their shoes and their rigidity can limit ankle flexion. Special foot straps for the trackers can allow them to be placed directly on the person's foot for better ankle freedom.

Similar tests were performed with two users suffering from neurological pathologies: User 7 with diabetic neuropathy and User 8 with stroke. They followed the same testing procedure, but tested the non-VR version of the system. Since this version does not allow the automatic evaluation of the classical training exercises (as it does not use the HTC Vive sensors), we measured them only manually using the goniometer and focused on the number of repetitions performed by the users. They performed the exercises in a sitting position, therefore the hip abduction exercise was skipped (must be performed standing up or laying on one side). One of the phenomena observed during the tests was that the users were obviously ambitioned to perform the exercises as well as possible—User 7 had difficulties in performing day-to-day tasks and had severe hand trembling when picking up objects or lifting them. However, they were concentrating better when performing the exercises of the system and obtained better results than expected for both upper limb (Table 10) and lower limb exercises (Table 11). For the elbow supination/pronation exercise, we displayed the results of the supination/pronation for the right arm on one row, and those of the left arm on the next row.

Table 10. Upper limb mobility statistics for each non-healthy user. All corresponding scores for the ROM are mentioned in **Table A1**.

| User ID | Shoulder Flexion | | | Shoulder Abduction | | | Elbow Flexion | | | Elbow Supination/Pronation | | | Arm Pushing | | | Fist Extension | | |
|---|---|---|---|---|---|---|---|---|---|---|---|---|---|---|---|---|---|---|
| | Angle Right/Left | ROM Right/Left | t(s) Right + Left | Angle Right/Left | ROM Right/Left | t(s) Right + Left | Angle Right/Left | ROM Right/Left | t(s) Right + Left | Angle Right/Left | ROM Right/Left | t(s) Right + Left | Angle Right/Left | ROM Right/Left | t(s) Right + Left | Angle Right/Left | ROM Right/Left | t(s) Right + Left |
| User 7 | 46 / 34 | 2 / 2 | 160 | 54 / 44 | 3 / 3 | 145 | 127 / 136 | 5 / 5 | 252 | 72/59 / 73/44 | 4/4 / 4/3 | 165 | 25 / 24 | 2 / 2 | 195 | 61 / 73 | 4 / 5 | 200 |
| User 8 | 72 / 76 | 4 / 4 | 125 | 79 / 78 | 4 / 4 | 165 | 129 / 142 | 5 / 5 | 250 | 89/83 / 90/88 | 5/5 / 5/5 | 160 | 40 / 51 | 3 / 4 | 175 | 55 / 52 | 4 / 4 | 145 |

**Table 11.** Lower limb mobility statistics for each healthy user and average results across all healthy users. All corresponding scores for the ROM are mentioned in **Table A1**.

| User id | Hip Flexion | | | Knee Flexion | | | Ankle Flexion | | |
|---|---|---|---|---|---|---|---|---|---|
| | Avg. Angle Right/left | ROM Right/Left | t(s) Right + Left | Avg. Angle Right/Left | ROM Right/Left | t(s) Right + Left | Avg. Angle Right/Left | ROM Right/Left | t(s) Right + Left |
| User 7 | 47<br>52 | 4<br>4 | 160 | 10<br>15 | 1<br>2 | 210 | 16<br>22 | 3<br>4 | 120 |
| User 8 | 71<br>72 | 4<br>4 | 125 | 30<br>30 | 3<br>3 | 125 | 25<br>21 | 4<br>4 | 155 |

We observe that in the case of non-healthy users, the numerical evaluation is better for unitary exercises (e.g., fist extension) rather than for complex ones (e.g., arm pushing). For instance, arm pushing and fist extension focus on the same joint evaluation (fist joint), but in the case of arm pushing, the movement is more complex, requiring more movements performed simultaneously (arm stretching, elbow extension) and patients tend to concentrate on performing the complex movement correctly rather than having a high fist joint angle. This can occur because of the deterioration of neurological functions with ageing which can affect concentration and multitasking capabilities [87]. In rehabilitation procedures, the time of execution is less important than the action of performing with success all the repetitions of an exercise which can further lead to the recovery of a lost function. Therefore, each one of the users was asked to perform the exercises at their own pace, with the difference in the execution time (Tables 8–11) being mostly related to the age or sex differences (e.g., younger people have usually better joint mobility and were able to perform the exercises faster, males have usually poorer lower limb mobility than females and therefore executed lower limb exercises slower). These aspects are, however, not very relevant for our system, as there is no need to oblige the users to perform the exercises at a certain speed, and they should concentrate on performing them correctly and with the required number of repetitions.

For determining the accuracy related to the system's evaluation of the healthy user's movements in the training tutorial, the real angles of joints were measured using a goniometer mobile application [88] and compared with the values registered by the system. One measurement was performed with the goniometer at the beginning of each exercise, with each one of the users, and compared with the same trial as registered by the system. The accuracy for each measurement and the average accuracy for each exercise can be found in Table 12 for the upper limb and Table 13 for the lower limb.

**Table 12.** Upper limb exercises accuracy—goniometer and system comparison.

| User ID | Shoulder Flexion | | | Shoulder Abduction | | | Elbow Flexion | | | Elbow Supination | | | Arm Pushing | | | Fist Extension | | |
|---|---|---|---|---|---|---|---|---|---|---|---|---|---|---|---|---|---|---|
| | Angle (gon.) | Angle (sys.) | Acc. (%) | Angle (gon.) | Angle (sys.) | Acc. (%) | Angle (gon.) | Angle (sys.) | Acc. (%) | Angle (gon.) | Angle (sys.) | Acc. (%) | Angle (gon.) | Angle (sys.) | Acc. (%) | Angle (gon.) | Angle (sys.) | Acc. (%) |
| User 1 | 86 | 94 | 90.7 | 83 | 85 | 97.6 | 130 | 123 | 94.61 | 37 | 43 | 83.78 | 55 | 57 | 96.36 | 57 | 57 | 100 |
| User 2 | 89 | 92 | 96.63 | 92 | 95 | 96.73 | 135 | 141 | 95.55 | 84 | 89 | 94.04 | 60 | 59 | 98.33 | 69 | 62 | 89.85 |
| User 3 | 86 | 86 | 100 | 88 | 91 | 96.59 | 115 | 113 | 98.26 | 58 | 61 | 94.82 | 46 | 47 | 97.82 | 30 | 32 | 93.33 |
| User 4 | 98 | 105 | 92.86 | 90 | 92 | 97.77 | 118 | 114 | 96.61 | 75 | 74 | 98.67 | 70 | 65 | 92.85 | 60 | 59 | 98.33 |
| User 5 | 92 | 96 | 95.66 | 81 | 88 | 91.35 | 137 | 130 | 94.89 | 81 | 83 | 97.53 | 68 | 62 | 91.17 | 61 | 54 | 88.52 |
| User 6 | 84 | 85 | 98.89 | 81 | 80 | 98.76 | 113 | 116 | 97.34 | 83 | 82 | 98.79 | 34 | 34 | 100 | 54 | 55 | 98.14 |
| Avg. Acc. (%) | | 95.79 | | | 96.46 | | | 96.21 | | | 94.60 | | | 96.08 | | | 94.70 | |

**Table 13.** Lower limb exercises accuracy—goniometer and system comparison.

| User ID * | Hip Flexion | | | Hip Abduction | | | Knee Flexion | | | Ankle Flexion | | |
|---|---|---|---|---|---|---|---|---|---|---|---|---|
| | Angle (gon.) | Angle (sys.) | Acc. (%) | Angle (gon.) | Angle (sys.) | Acc. (%) | Angle (gon.) | Angle (sys.) | Acc. (%) | Angle (gon.) | Angle (sys.) | Acc. (%) |
| User 1 | 80 | 81 | 98.75 | 60 | 58 | 96.66 | 102 | 110 | 92.15 | 35 | 31 | 88.57 |
| User 2 | 72 | 73 | 98.61 | 40 | 44 | 90 | 66 | 64 | 96.96 | 32 | 31 | 96.87 |
| User 3 | 92 | 90 | 97.83 | 60 | 62 | 96.66 | 125 | 128 | 97.6 | 32 | 31 | 96.87 |
| User 5 | 67 | 68 | 98.5 | 47 | 52 | 89.36 | 80 | 90 | 87.5 | 33 | 35 | 93.94 |
| User 6 | 82 | 82 | 100 | 54 | 54 | 100 | 80 | 78 | 97.5 | 25 | 26 | 96 |
| Avg. Acc. (%) | | 98.74 | | | 94.54 | | | 94.34 | | | 94.45 | |

The average accuracy obtained across all upper and lower limb exercises was of 95.59%, showing great potential for future developments and use with neurorehabilitation patients.

For phase 2 of the testing procedure, we analyzed the performance obtained by each healthy user based on the score from each one of the trials, for each game (Table 14). For consistency reasons, we established a similar five-point Likert scale, with each grade described according to each game and difficulty level in Appendix A, Table A2. Hit targets is the only exercise where we kept the adaptive algorithm to see the highest level of difficulty reached by a user; all other games included three individual trials of one minute each. For personal health reasons, User 4 only executed the games which targeted the upper limb. A detailed score for each user can be consulted in the Supplementary Materials (S2_CompleteUserTestingData, Tables Table_User1 – Table_User8).

**Table 14.** Games scores for each user. All corresponding scores are mentioned in **Table A2**.

| User ID | Hit Targets 3 Min Adaptively | Ball Directing Trial 1 | Trial 2 | Trial 3 | Whack-A-Mole Easy | Med. | Hard | Boxing Ring 1 | Ring 2 | Bag | Football Pen. 1 | Pen. 2 | Free kick |
|---|---|---|---|---|---|---|---|---|---|---|---|---|---|
| User 1 | 3 (reached medium level) | 5 | 5 | 5 | 5 | 4 | 4 | 5 | 5 | 5 | 4 | 5 | 4 |
| User 2 | 4 (almost completed medium) | 3 | 5 | 5 | 3 | 4 | 4 | 5 | 4 | 5 | 3 | 3 | 3 |
| User 3 | 5 (reached and completed difficult) | 5 | 5 | 5 | 4 | 5 | 4 | 4 | 4 | 3 | 4 | 4 | 4 |
| User 4 * | 2 (2 easy trials started but not completed) | 1 | 2 | 4 | 2 | 2 | 3 | 4 | 4 | 4 | - | - | - |
| User 5 | 2 (2 easy trials started but not completed) | 4 | 3 | 5 | 5 | 5 | 3 | 5 | 5 | 5 | 3 | 3 | 3 |
| User 6 | 5 (completed multiple difficult levels) | 3 | 4 | 5 | 4 | 4 | 4 | 5 | 5 | 5 | 3 | 4 | 4 |
| Avg. | | 3.5 | 4 | 4.83 | 3.83 | 4 | 3.67 | 4.67 | 4.5 | 4.5 | 3.4 | 3.8 | 3.6 |

* User 4 performed only upper limb games for personal health reasons.

The games which had the best results among all users and all trials were boxing, ball directing and whack-a-mole. Boxing and whack-a-mole required more basic movements (arm movement when holding the controller, without interacting with its input) and were easier to understand and use by all users, regardless of their level of experience with technology. The "hit targets" game required a more complex movement (clenching fist to press the controller's grip buttons for picking up the ball, raising arm, coordinating the grip release moment to launch the ball in the right direction) and therefore required more accommodation time. The football game would benefit from the personalization of the user's avatar based on their height and lower limb dimensions.

In Figure 19, the column chart shows for each game and each trial the users' performance, in terms of their obtained grades. Trial 3 of the Ball directing game had the best results (five users with excellent performance, one user with good performance), while the first penalty trial of the football game had the worst results (three users with fair performance and two users with good performance). All games have a 100% accuracy (recorded data compared with self-assessment or observation from the supervisor/trainer—e.g., number of collisions with the punching bag, number of cans hit).

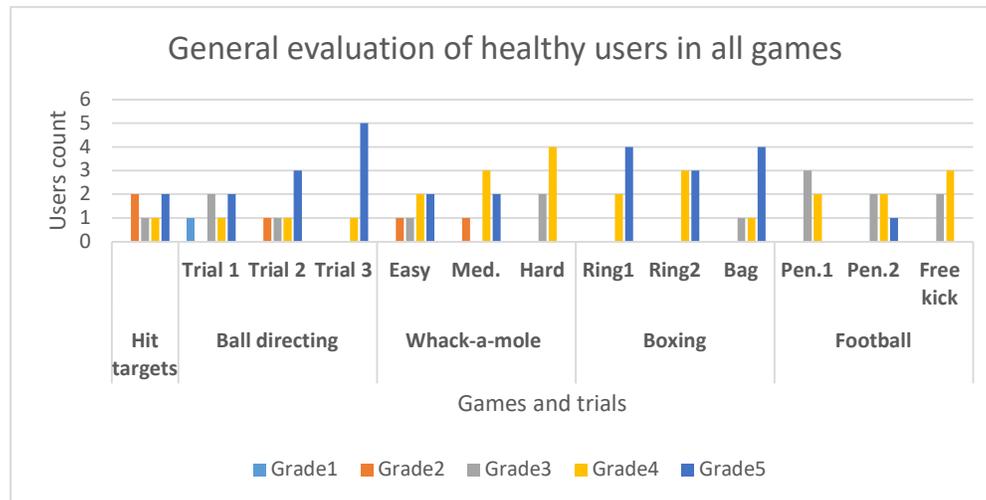

**Figure 19.** Number of users which obtained a certain grade performance (1–5 Likert scale) for all games and trials.

Virtual reality and technology experience have no impact on the users' accuracy in the classical training exercises. User 6, despite their lack of VR experience and only fair technology experience, performed good and very good in all training exercises, both classical and gamified. User 6 had great performance in all gamified exercises. If we compare the performances in relation to the user's age, we can thus observe that they are not correlated, as people of all ages can perform well in all games. It should be noted that User 6 has a history of playing professional handball in their youth.

In the case of non-VR training with people affected by neurological disorders, the Myo armband was not able to detect with a high accuracy the "clenching fist" gesture as input for game actions (e.g., picking up a ball). This caused frustration, as the user was able to pick up the ball occasionally, but Myo was detecting their fist release earlier, resulting in the ball dropping without throwing it towards the targets. A redesign is thus needed for games which require this gesture (i.e., hitting targets, ball directing) in the case of the non-VR system setting. We obtained better gesture detection in the case of hand waving which can be used as a replacement, or we can take into consideration the value provided by the peak to peak amplitude levels detected from the EMG sensors for upper forearm muscles. The boxing and whack-a-mole games were successfully performed by Users 7 and 8, as they required only rotation and direction input from the Myo armband. Coordination and lower reaction times in the case of non-healthy subjects show the importance of adapting the exercises to their pathology magnitude, as standard configurations lead to low accuracies (whack a mole—16.9% average accuracy for User 7, 27.13% average accuracy for User 8). On the other hand, boxing seems to be an exercise suitable and pleasant for all users, regardless of their health condition—Users 7 and 8 obtained mostly good and very good boxing scores according to Table A2 (Annex 1) classification.

Heart rate was monitored continuously during the training procedure in order to analyze its impact on the person's effort levels. Table 15 shows the initial pulse of each user (when resting, in phase 0), the maximum pulse reached during training, the medium pulse across all training exercises as well as effort levels (relaxed, low or intense), as percentages of the training session.

Table 15. Heart rate values and effort levels.

| User ID | Heart Rate | | | Effort Levels (% of Entire Training Session) | | |
| --- | --- | --- | --- | --- | --- | --- |
| | Initial Value | Max. Value | Medium Value | Relaxed | Low | Intense |
| User 1 | 100 | 134 | 105 | 19.85% | 58.78% | 21.37% |
| User 2 | 66 | 118 | 85 | 90.63% | 9.37% | 0.00% |
| User 3 | 84 | 112 | 92 | 70.59% | 22.35% | 7.06% |
| User 4 * | 83 | 100 | 83 | 98.96% | 1.04% | 0.00% |
| User 5 | 72 | 121 | 86 | 72.39% | 25.77% | 1.84% |
| User 6 | 82 | 121 | 97 | 38.75% | 55.81% | 5.44% |
| User 7 * | 68 | 81 | 70 | 100% | 0.00% | 0.00% |
| User 8 * | 72 | 95 | 66 | 100% | 0.00% | 0.00& |

* User 4, User 7 and User 8 performed the tests sitting.

For Users 4, 7 and 8, we observe that the effort level was the lowest, with more than 98% of the training being performed at relaxed heart rate levels. As these users were performing the exercises while sitting, yet still obtaining good results during training, we can conclude that the system can be successfully used by people that cannot perform at high effort because of their health condition.

For all users, we collected EMG data sent by the Myo gesture control armband via Bluetooth at 200 Hz (Supplementary Materials S4_EMG). All data were collected on the eight individual channels corresponding to the eight sensors of Myo (Figure 20). Data were grouped based on the exercises performed by each user, varying between 2500 and 16,000 EMG signal output values per exercise. The signal was normalized by default in the -128, 127 range and as shown by Nasri et al., it can be further interpreted by using the window method to select relevant data regions, classified by the use of neural networks to identify hand movements and gestures [89]. As underlined by Slutter et al., the action of "squeezing" tensions the upper muscles in the forearm corresponds to the EMG0, EMG1 and EMG7 sensors [90]. As the exercise was performed while "squeezing" the HTC Vive controller, higher amplitude (peak-to-peak) values can be observed for these three specific sensors in Figure 20.

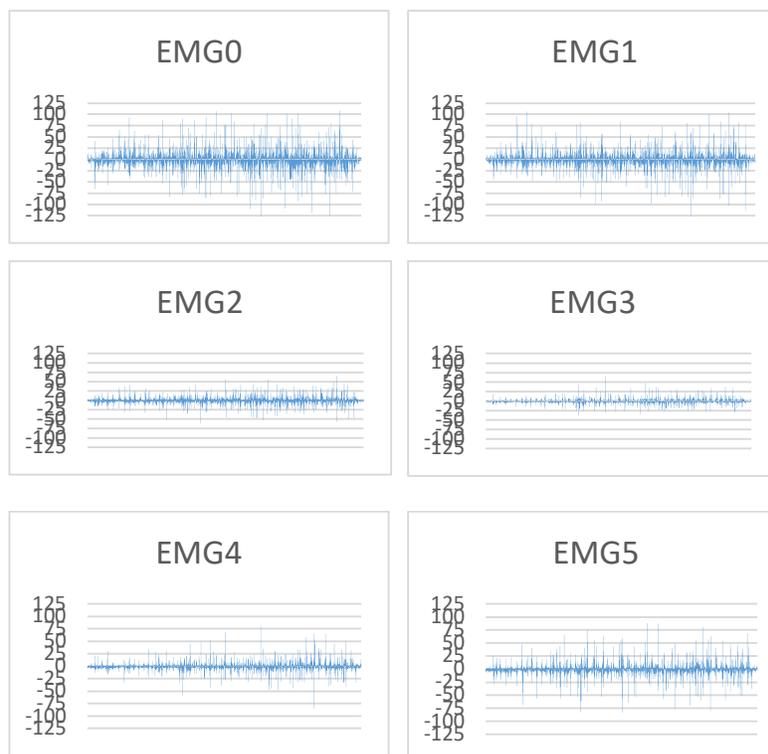

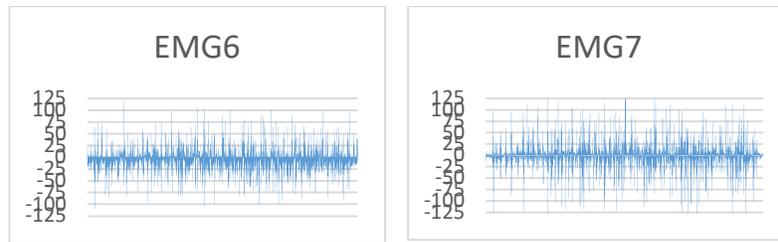

**Figure 20.** EMG signals for the 8 sensors of the Myo gesture control armband collected for User 6 during a whack-a-mole game.

*5.3. Feedback and Future Improvements*

In our testing procedure, we involved either physicians or auxiliary medical personnel from different medical specialties as test subjects, as well as people affected by neurological disorders. All tests were performed under the surveillance of a therapist which was part of the development team. All users appreciated the system and liked the graphics, scenes and ambience. When asked to choose where they wanted to perform the initial tutorial exercises, all users preferred the nature setting as opposed to the rehabilitation center one.

The gamified exercises were appreciated as useful, interesting and entertaining, and they animated the users' competitive spirit. One of the physicians appreciated that the games were essential for rehabilitation, as psychological limitations are overcome by the desire of self-improvement and movements remain restrained only by the patient's physical limits. Complex movements are being trained and the user is pushed toward progress and easier rehabilitation. Suggestions included some changes to the classical training exercises procedure—the trainer should exemplify the movements to the user and they must be executed at the same pace as they are being shown, not at each user's desired pace. This should also be configured by the therapist based on the patient's condition. People affected by neurological disorders appreciated that daily training including all exercises would be tiring and each day should include only a few exercises, changing from one day to another for diversity reasons. The test subjects would love to see more games developed for training complex movements.

Related to the use of a virtual reality system, even though most test subjects had no prior experience with this technology, none of them signaled any side effects. One of the physicians was expecting to feel dizzy and have an unpleasant experience prior to the use of the system but was pleasantly surprised as they were getting more and more used to the system from one trial to another. Future tests should be conducted with patients suffering from neurological disorders, including tests performed with the same subject for several days in a row, to observe progress and the long-term effects of the training process.

Overall, the good accuracy of 95.59% shows great potential, yet changes can be made for ameliorating it. We found no current studies which evaluated rehabilitation exercises performed by a user based on the most recent tracking technology used for the HTC Vive Cosmos Elite system, yet Jost et al. showed that the HTC Vive system could offer a higher accuracy in both position and rotation for a robotic arm's joints [48]. For improving our accuracy, future tests should be performed using trackers instead of controllers, for eliminating any differences between the tracking devices' initial and final position and orientation relative to the person's hand. Trackers are also very light (89 g each [91]) and do not have to be picked up or held, as they are fixed using straps on the person's fist joint or foot, respectively. Furthermore, the AnimationRigging Unity package used for developing the inverse kinematics-based animations of our real-time avatar is still in the beta stage of development and its performance should be analyzed after its final launch. In addition, goniometer angles measured by using mobile applications depend on the mobile phone's gyroscope precision. Studies show that their error can be up to 2.05° without performing a full calibration [92], so a physical goniometer should be used in future tests.

One of the biggest improvements which can also have a great impact on the accuracy is related to the personalization of the 3D avatar according to the user's parameters. Currently, custom changes can be made to fit the trackers on the user's hands/feet based on position and rotation offsets, but the 3D model's body parts cannot be scaled according to the user's height and limbs. This aspect is essential, as inverse kinematics animations act differently if body segments (i.e., longer arms, forearms) have different lengths. For adapting the use of case scenarios to each custom user, reference poses should be recorded for each patient (e.g., user relaxed, with arms to the side), not use the doctor's standard recorded poses as comparison basis. Improved 3D avatars can also reduce the limitation of movements, which is sometimes influenced by the model's mesh (e.g., forearm flexion cannot exceed a certain threshold if the mesh is not deformable to imitate the human skin and will produce unnatural effects).

## 6. Conclusions

In this paper, we presented INREX-VR, a virtual reality system using immersive exercises for neurorehabilitation. People suffering from neurological disorders see their lives changing dramatically—from day-to-day activities to occupation and social relationships, the impact of the disease over their quality of life is considerable. We took the evolution of sensors-based hardware devices as a good opportunity for capturing training exercises' movements and performing rehabilitation games for improving people's QoL through self-training. With the help of a great variety of motion detection and biological sensors (HTC Vive system, Myo gesture control armband including EMG), the system was suitable for both patients' rehabilitation process and for medical personnel training. The main goal was to create a flexible, extensible system, containing gamified training exercises for a wide range of neurological conditions, which can improve the quality of life of the affected persons. All data provided by the sensors and results of the training exercises are centralized in order to be evaluated or configured by specialists using a mobile application. Besides the imitation of classical exercises performed by virtual therapists, the system also includes gamified settings, since repeating classical exercises can often decrease motivation as they lack entertainment and competition. Games include an adaptive learning algorithm which can increase the difficulty levels in real time based on the user's performance. After discussions with rehabilitation specialists, we confirmed that the system showed potential of use for the training of the medical personnel involved in the rehabilitation field.

As presented by Corbetta et al., virtual reality systems can often prove to have better results for neurorehabilitation than standard therapy for walking speed, balance and mobility [93]. Our results are encouraging, yet future tests should also include neurorehabilitation patients for validating the system with impaired persons. Validity threats include the improper configuration and misuse of the system (e.g., base stations not placed at a recommended height and angle, controllers moved independently for "deceiving" the evaluation process). Misconfiguring the HTC Vive system and not updating the SteamVR software regularly can lead to tracking issues or lost tracking (e.g., one controller disappears in the middle of the training process). These issues can be resolved by providing proper instructions about the system's use and ensuring technical assistance for configuration or solving tracking issues.

In the future, we intend to improve the accuracy of movement evaluation, and plan to develop the adaptive learning algorithm so that it is also applied on the classical training exercises, test the tele-medicine module with patients and include quizzes for assessing the knowledge of the medical personnel which use the application for their own training. As the Myo armband is not available anymore on the market, alternatives are needed for reproducing the project or distributing it on a larger scale. One of the alternatives is MyoWare Muscle Sensor, provided by Sparkfun. It is an EMG sensor powered by an Arduino board, which also allows the attachment of biomedical pads [94]. The use of eight such sensors will reproduce the output provided by the Myo armband in terms of EMG data, while the Arduino can be easily integrated into Unity with the use of a plugin [95]. Therefore, all data from the EMG can be used as input for the rehabilitation games, using peak-to-peak values from the sensors instead of the predefined gestures provided by Myo. This technology has been used in research projects for assessing the physiological state of a person, in connection with Unity3D and the virtual reality HTC Vive headset [96]. Further research is also needed to address people with stereo blindness (who are not capable of synchronizing the two optical streams), for whom we take into consideration disabling 3D stereoscopic vision and rendering the scene only on the left/right eye. Furthermore, we plan to investigate the possibility of developing a light version of the system which could be used with low-cost mobile-based VR headsets for upper limb rehabilitation and include neural networks for improving the movements' automatic evaluation. We consider that INREX-VR opens new opportunities in terms of using technology advancement in the field of neurorehabilitation, not as a replacement of a real-life therapist, but as a complementary tool.

**Supplementary Materials:** The following are available online at http://www.mdpi.com/1424-8220/20/21/6045/s1, S1: Patient-consent-form, S2: Complete User Testing Data, S3: Recording Animations, S4: EMG.

**Author Contributions:** Conceptualization, I.-C.S. and A.M.; methodology, I.-C.S.; software, I.-C.S.; validation, I.-C.S, G.-P.P., M.G.R., M.-I.D..; formal analysis, I.-C.S.; investigation, I.-C.S., G.-P.P., M.G.R.; resources, A.M.; data curation, I.-C.S.; writing—original draft preparation, I.-C.S., G.-P.P., M.G.R.; writing—review and editing, M.-I.D., F.M., A.M.; visualization, all authors.; supervision, F.M., M.-I.D.; project administration, F.M., A.M., M.-I.D. and I.-C.S. All authors have read and agreed to the published version of the manuscript.

**Funding:** This research received no external funding.

**Acknowledgments:** The authors would like to thank all participants involved in the testing procedure. This work was supported with material and logistic aid by the Competitiveness Operational Program 2014–2020, Action 1.1.3: Creating synergies with RDI actions of the EU's HORIZON 2020 framework program and other international RDI programs, MySMIS Code 108792, Acronym project "UPB4H", financed by contract: 250/11.05.2020.

**Conflicts of Interest:** The authors declare no conflict of interest.

## Abbreviations

The following abbreviations are used in this manuscript:

| | |
|---|---|
| BCI | Brain–computer interface |
| DOI | Digital object identifier |
| EEG | Electroencephalogram |
| EMG | Electromyography |
| ICF | International Classification of Functioning, Disability and Health |
| IK | Inverse kinematics |
| INREX-VR | Immersive Neurorehabilitation Exercises Using Virtual Reality |
| ISI | International Scientific Index |
| QoL | Quality of life |
| PD | Parkinson's disease |
| PRISMA | Preferred Reporting Items for Systematic Reviews and Meta Analyses |
| ROM | Range of motion |
| SDK | Software development kit |
| VR | Virtual reality |

# Appendix A

**Table A1.** ROM degree Likert scale for each training exercise.

| Exercise | Excellent (5) | Good (4) | Fair (3) | Poor (2) | Very Poor (1) |
|---|---|---|---|---|---|
| Upper limb | | | | | |
| Shoulder flexion | >80° | 65°–80° | 50°–65° | 30°–50° | <30° |
| Shoulder abduction | >80° | 60°–80° | 40°–60° | 25°–40° | <25° |
| Elbow flexion | >100° | 80°–100° | 40°–80° | 20°–40° | <20° |
| Elbow supination | 75°–90° | 50°–75° | 35°–50° | 20°–35° | <20° |
| Arm pushing | >60° | 45°–60° | 30°–45° | 15°–30° | <15° |
| Fist extension | >70° | 50°–70° | 35°–50° | 20°–35° | <20° |
| Lower limb | | | | | |
| Hip flexion | >75° | 45°–75° | 30°–45° | 15°–30° | <15° |
| Hip abduction | >45° | 30°–45° | 20°–30° | 10°–20° | <10° |
| Knee flexion | >90° | 50°–90° | 30°–50° | 15°–30° | <15° |
| Ankle flexion | >35° | 20°–35° | 10°–20° | 5°–10° | <5° |

**Table A2.** Score-based Likert scale for each gamified exercise.

| Game | Excellent (5) | Good (4) | Fair (3) | Poor (2) | Very Poor (1) |
|---|---|---|---|---|---|
| Upper limb | | | | | |
| Hit targets (cans) | Reaching level "difficult" where at least one can is hit | All easy levels completed and half of the targets from a medium level are hit | All easy levels completed and reached medium | 2–3 cans hit in one easy level | 0–1 can hit in one easy level |
| Ball directing | >700 points | 500–700 points | 300–500 points | 100–300 points | <100 points |
| Whack-a-mole—easy | >95% hit accuracy | 75–95% hit accuracy | 55–75% hit accuracy | 35–55% hit accuracy | <35% hit accuracy |
| Whack-a-mole—medium | >90% hit accuracy | 70–90% hit accuracy | 50–70% hit accuracy | 30–50% hit accuracy | <30% hit accuracy |
| Whack-a-mole—hard | >85% hit accuracy | 65–85% hit accuracy | 45–65% hit accuracy | 25–45% hit accuracy | <25% hit accuracy |
| Boxing—ring | >50 hits/each fist | 40–50 hits/each fist | 30–40 hits/each fist | 20–30 hits/each fist | <20 hits/each fist |
| Boxing—bag | >40 hits/each fist | 30–40 hits/each fist | 20–30 hits/each fist | 10–20 hits/each fist | <10 hits/each fist |
| Lower limb | | | | | |
| Football—penalty | >10 goals | 7–10 goals | 4–7 goals | 2–4 goals | <2 goals |
| Football—free kick | >8 goals | 5–8 goals | 2–5 goals | 1–2 goals | 0 goals |